\documentclass[useAMS,usenatbib,usegraphicx]{mn2e}
\usepackage{psfig}   
\usepackage{graphicx}
\usepackage{subfigure}

\title[Do binaries in clusters form in the same way as in the field?]{Do binaries in
  clusters form in the same way as in the field?}

\author[R.~J.~Parker, S.~P.~Goodwin, P.~Kroupa and M.~B.~N.~Kouwenhoven]{
  Richard J.~Parker$^1$\thanks{E-mail: r.parker@sheffield.ac.uk},
  Simon P.~Goodwin$^1$, Pavel Kroupa$^2$ and M.~B.~N.~Kouwenhoven$^1$ \vspace*{0.1cm}\\
   $^1$ Department of Physics and Astronomy, University of Sheffield,
    Sheffield, S3 7RH, UK\\
   $^2$ Argelander Institut f\"{u}r Astronomie, Universit\"{a}t Bonn,
    Auf dem H\"{u}gel 71, 53121 Bonn, Germany}

\begin{document}

\date{}
                             
\pagerange{\pageref{firstpage}--\pageref{lastpage}} \pubyear{2009}

\maketitle

\label{firstpage}

\begin{abstract}
We examine the dynamical destruction of binary systems in star clusters 
of different densities.  We find that at high densities 
($10^4-10^5$\,M$_\odot$~pc$^{-3}$) almost all binaries with separations $>~10^3$~AU 
are destroyed after a few crossing times. At low densities 
(${\cal O}(10^2)$\,M$_\odot$~pc$^{-3}$) many binaries with separations $>10^3$~AU are 
destroyed, and no binaries with separations $>10^4$~AU survive after a few crossing times.  
Therefore the binary separations in clusters can be used as a tracer of 
the dynamical age and past density of a cluster.

We argue that the central region of the Orion Nebula Cluster was $\sim 100$ times
denser in the past with a half-mass radius of only $0.1$ -- $0.2$~pc 
as (a) it is expanding, (b) it has very few binaries with separations $>10^3$~AU, and
(c) it is well-mixed and therefore {\em dynamically} old.  

We also examine the origin of the field binary population.  Binaries
with separations $<10^2$~AU are not significantly modified in any cluster, therefore at 
these separations the field reflects the sum of all star formation.  
Binaries with separations in the range $10^2$ -- $10^4$~AU are progressively more and more heavily 
affected by dynamical disruption in increasingly dense clusters.  
If most star formation is clustered, these binaries must be over-produced 
relative to the field.  Finally, no binary with a separation $>10^4$~AU can survive in 
{\em any} cluster and so must be produced by isolated star formation, but only if 
{\em  all} isolated star formation produces extremely wide binaries.
\end{abstract}


\section{Introduction}

Most stars in the field are observed to be in multiple
systems\footnote{\citet{Lada06} points out that the binary fraction of
M-dwarfs is low and so most stellar {\em systems} are single.
However, the fraction of stars (rather than systems) that are in
binaries could be more than half.}
\citep[e.g.][]{Duquennoy91,Fischer92}.   The binary fraction of young
stars appears to be even higher than that of the field
\citep[e.g.][and references therein]{Mathieu94, Patience02, Goodwin05,
Goodwin07}.   \citet{Kroupa01b} show that in dense clusters it is
very difficult to make significant numbers of binaries  from initially
single stars and that it is  impossible to widen an initially narrow
separation distribution to the  observed wide distribution. Therefore,
the field binary fractions and properties must in some way mirror the
outcome of star formation.  Thus, one of the most significant
constraints on models of star formation is that these models correctly
predict the numbers and properties of binary and multiple systems
\citep{Goodwin07}.

However, at least 50 per cent (probably 70 -- 90 per cent) of 
stars are thought to form in clusters \citep[e.g.][]{Lada03}.   It is
known that internal dynamical processes in clusters can rapidly and 
significantly affect the properties of binaries in clusters 
\citep*[e.g.][]{Heggie75, Hills75a, Hills75b, Kroupa95a, Kroupa95b, Kroupa99}.
The degree to which binary properties will be altered depends also on the density
and lifetime of the cluster in which the binaries are born.

The field binary population is thus a mixture of binaries from different density 
environments (which may or may not be the same initial population) which have been 
dynamically processed in different ways. Therefore, the outcome of (hydrodynamical)
star formation simulations should not, and cannot, be directly compared with the field 
population. This paper is the first in a series in which we attempt to reconstruct 
the initial binary population resulting from the early stages (Class 0/I) of 
star formation.  This paper will investigate the dynamical processing of binaries in 
clusters of different densities and examine the extent to which binaries of different 
separations are processed.  This will allow us to discuss the main aspects of attempting 
to model the origin of the field, and reconstruct the initial binary population.

In this paper, we investigate how a field-like initial binary
population is modified in clusters.  In Section~\ref{method}
we describe our initial conditions; we  present our results and discussion in 
Sections~\ref{results}, \ref{ONCoriginal} and \ref{discussion}, and we conclude in 
Section~\ref{conclude}.

\section{Method}
\label{method}

\subsection{Initial conditions}

The simulated star clusters have masses $\sim 10^2$ -- $10^3$\,M$_\odot$
and a range of half-mass radii of 0.1, 0.2, 0.4 and 0.8\,pc.  This
means we can simulate clusters with a range of densities from 
$\sim 50$\,M$_\odot$ pc$^{-3}$ to $\sim 10^5$\,M$_\odot$ pc$^{-3}$ covering almost
the complete range of probable initial cluster densities.  We 
summarise the properties of the clusters that we simulate in
  Table~\ref{table}.

For each set of initial conditions we run an ensemble of at least $10$
simulations which are identical apart from the random number seed used
to initialise the positions, masses and binary properties.

Our clusters are set-up as initially virialised Plummer spheres
\citep{Plummer11}  according to the prescription of
\citet*{Aarseth74}.  The Plummer sphere  provides the positions and
velocities of the centres of  mass of systems -- which may be single or
binary systems.  

To create a stellar system, the mass of the primary star is chosen 
randomly from a \citet{Kroupa02} IMF of the  form
\begin{equation}
 N(M)   \propto  \left\{ \begin{array}{ll} M^{-1.3} \hspace{0.4cm} m_0
  < M/{\rm M_\odot} < m_1   \,, \\ M^{-2.3} \hspace{0.4cm} m_1 <
  M/{\rm M_\odot} < m_2   \,,
\end{array} \right.
\end{equation}
where $m_0$ = 0.1\,M$_\odot$, $m_1$ = 0.5\,M$_\odot$, and  $m_2$ =
50\,M$_\odot$. For simplicity we do not include brown dwarfs (BDs)  in
our simulations. The effect of dynamical processing on BDs in
clusters will be studied in a future paper (for the results of recent
observations of BDs in Orion see \citealt{Maxted08}; and for existing
theoretical work see \citealt{Kroupa03} and \citealt{Thies08}).

We then assign a secondary to the system depending on the binary
fraction associated with the primary mass.

For a field-like binary fraction we divide primaries into four
groups. Primary masses in the range 0.08~$\leq M/{\rm M}_\odot~<$~0.47 
are M-dwarfs, with a binary fraction of 0.42
\citep{Fischer92}\footnote{We note that the M-dwarf binary
    fraction is highly uncertain.  The results of \citet{Fischer92} 
    are probably only appropriate for stars with mass $>0.3$\,M$_\odot$,
    and the binarity and separation distributions below this mass may
    be quite different (see especially \citealt{Lada06} and also \citealt{Maxted08}).}.
K-dwarfs have masses in  the range
0.47~$\leq~M/{\rm M}_\odot$~$<$~0.84 and binary fraction of 0.45
\citep{Mayor92} and G-dwarfs have masses from 0.84~$\leq~M/{\rm
M}_\odot~<$~1.2  with a binary fraction of 0.57
\citep{Duquennoy91}. All stars more massive than  1.2\,M$_\odot$ are
grouped together and assigned a binary fraction of unity, as  massive
stars have a much larger binary fraction than low-mass stars
\citep[e.g.][and references
therein]{Abt90,Mason98,Kouwenhoven05,Kouwenhoven07,Pfalzner07}.

Clusters with an initial binary fraction of unity for {\em all} stars
are also created in order to test the hypothesis that all  stars form
in binary systems and that single stars are solely the result  of the
dynamical processing of binaries \citep{Kroupa95a,Kroupa95b,Goodwin05}.

\begin{table}
\caption[bf]{A summary of the different cluster properties adopted for the simulations.
The values in the columns are: the number of stars in each cluster ($N_{\rm stars}$), 
the typical mass of this cluster ($M_{\rm cluster}$), the initial half-mass radius of 
the cluster ($r_{\rm 1/2}$), the crossing time of the cluster ($t_{\rm cross}$) and the 
initial binary fraction in the cluster ($f_{\rm bin}$).}
\begin{center}
\begin{tabular}{|c|c|c|c|c|}
\hline 
$N_{\rm stars}$ & $M_{\rm cluster}$  & $r_{\rm 1/2}$ & $t_{\rm cross}$ & $f_{\rm bin}$  \\
\hline
1500 & $\sim 10^3$\,M$_\odot$ & 0.1\,pc & $\sim$ 0.02\,Myr & 100\,per cent \\
1500 & $\sim 10^3$\,M$_\odot$ & 0.1\,pc & $\sim$ 0.02\,Myr & field-like \\
1500 & $\sim 10^3$\,M$_\odot$ & 0.2\,pc & $\sim$ 0.05\,Myr & 100\,per cent \\
1500 & $\sim 10^3$\,M$_\odot$ & 0.2\,pc & $\sim$ 0.05\,Myr & field-like \\
1500 & $\sim 10^3$\,M$_\odot$ & 0.4\,pc & $\sim$ 0.1\,Myr & field-like \\
1500 & $\sim 10^3$\,M$_\odot$ & 0.8\,pc & $\sim$ 0.4\,Myr & 100\,per cent \\
1500 & $\sim 10^3$\,M$_\odot$ & 0.8\,pc & $\sim$ 0.4\,Myr & field-like \\
\hline 
100 & $\sim 10^2$\,M$_\odot$ & 0.4\,pc & $\sim$ 0.4\,Myr & 100\,per cent \\
100 & $\sim 10^2$\,M$_\odot$ & 0.4\,pc & $\sim$ 0.4\,Myr & field-like \\
\hline
\end{tabular}
\end{center}
\label{table}
\end{table}

\subsection{Binary properties}

Secondary masses are drawn from a flat mass ratio distribution with
the constraint that if the companion mass is $<0.1$\,M$_\odot$ it is
reselected.  This maintains the underlying binary fraction, but biases
the mass ratios of low-mass systems towards unity
\citep[see][]{Kouwenhoven09a}.  We note that this means that
we do not recover a Kroupa IMF despite our primaries being drawn from
this distribution.

Eccentricities of binary stars are drawn from a thermal eccentricity
distribution \citep{Heggie75,Kroupa95a,Kroupa08} of the form
\begin{equation}
f_e(e) = 2e.
\end{equation}

The generating function for orbital periods are the log-normal
distributions observed by \citet{Duquennoy91} and \citet{Fischer92} of
the form
\begin{equation}
f\left({\rm log}P\right) = C{\rm exp}\left \{ \frac{-{({\rm log}P -
\overline{{\rm log}P})}^2}{2\sigma^2_{{\rm log}P}}\right \},
\end{equation}
where $\overline{{\rm log}P} = 4.8$, $\sigma_{{\rm log}P} = 2.3$ and
$P$ is  in days. The periods are then converted to semi-major
axes. Binary systems  with small semi-major axes but large
eccentricities are expected to undergo  tidal circularisation, as
observed in \citeauthor{Duquennoy91}'s \citeyearpar{Duquennoy91}
sample.  We  include the effects of tidal circularisation in our
simulations  by utilising \citeauthor{Kroupa95b}'s
\citeyearpar{Kroupa95b} `Eigenevolution' process.  However, we
  note that the systems that are circularised by eigenevolution are so
  hard as to be generally unaffected by dynamical interactions and so
  the details of eigenevolution are unimportant (we also ran
  simulations without any eigenevolution and also with a simplified
  circularisation mechanism and found no significant differences).

By combining the primary and secondary masses of the binaries with
their  semi-major axes and eccentricities, the relative velocity and
radial  components of the stars in each system are determined. These
are then  placed at the centre of mass and centre of velocity for each
system in the  Plummer sphere.

Simulations are run using the \texttt{kira} integrator in Starlab
\citep[e.g.][and references therein]{Zwart99,Zwart01} and evolved  for
10\,Myr.  Note that we do not include stellar evolution; we have
checked that this makes very little difference to the outcome of the
simulations as there are few high-mass stars that evolve on such a
short timescale (and, as we shall see, most binary evolution occurs in
the first Myr, before the massive stars have evolved).

\subsection{Finding bound binary systems}

We determine whether a star is in a binary system using the following
method\footnote{Our algorithm almost exactly
reproduces the results of the independent binary finding algorithm of
\citet[][in prep.]{Kouwenhoven09b}.}. For each star, the distances to
its ten nearest neighbours are determined. These distances are then
used to find the average distance between stars locally.  We then
determine the identities of the nearest and second nearest neighbours
of each star.  If two stars are mutual nearest neighbours with a
separation of less than a quarter of the local average distance
between stars they are a potential binary\footnote{This method is also
able to find higher-order systems by examining the relationships
between sets of three or more stars.  For example a triple is a system
in which three stars are mutual nearest and second nearest
neighbours.}.  Numerical experiments have shown that it is extremely
rare to find a bound system with a separation greater than a quarter
of the local average separation. They are a {\em true} binary if they
also have a significantly negative relative energy, and they are an
{\em observational} binary if they have positive energy (i.e. an
observer may think the system is a binary when it is just two stars
passing).  Clearly in our definition an observational binary can
  be found in either 2D or 3D, however only 2D observational binaries
  are of interest as they represent potential mistakes in actual
  observations.  However, for the rest of this paper we ignore
observational binaries as they are found to be extremely rare.

%                              RESULTS

\section{Clusters of different densities}
\label{results}

In this section we will examine the evolution of the binary
populations in clusters of different densities.

Firstly, we will consider dense clusters with half-mass radii of $0.1$
-- $0.2$ pc and masses of $\sim 750$\,M$_\odot$ ($N_{\rm stars}~\sim~1500$) with
densities of $10^4-10^5$\,M$_\odot$ pc$^{-3}$.  Then we will
consider low-density clusters with large half-mass radii, or low
numbers of stars, with densities of ${\cal O}(10^2)$\,M$_\odot$~pc$^{-3}$.  

These two density regimes cover the whole range of initial cluster
densities in probably equal proportions.  If the cluster mass function
is $N(M) \propto M^{-2}$ \citep{Lada03} then each equal
logarithmic mass bin contributes the same {\em mass} of stars to the
field.  Taking the range of cluster masses to be $10^1$ -- $10^5$\,M$_\odot$, 
then an equal mass of stars is produced by clusters larger
than Orion ($\sim 10^3$\,M$_\odot$) to the mass produced by 
clusters smaller than Orion.  The higher densities are probably fairly
representative of larger clusters, and the lower densities of smaller
clusters (we will return to this in Section~\ref{ONCoriginal} where we examine the
initial conditions of Orion).

We might well expect the evolution of the binary populations in
different density clusters to be very different as the 
position of the hard-soft boundary, $a_{\rm h-s}$,   will depend
on the number of systems $N_{\rm sys}$ in, and the  half-mass radius
$r_{1/2}$, of the cluster \citep[following][]{Binney87},
\begin{equation}
a_{\rm h-s} \propto \frac{r_{1/2}}{N_{\rm sys}}.
\label{eqn:hs}
\end{equation}
Thus the hard-soft boundary will shift by a factor of nearly $100$
between the least and most dense clusters.

\subsection{Dense clusters}

\subsubsection{The evolution of the binary fraction}

\begin{figure*}
  \begin{center}
\setlength{\subfigcapskip}{10pt} \subfigure[Field binary
    fraction.]{\label{fmult-a}\rotatebox{270}{\includegraphics
    [scale=0.4]{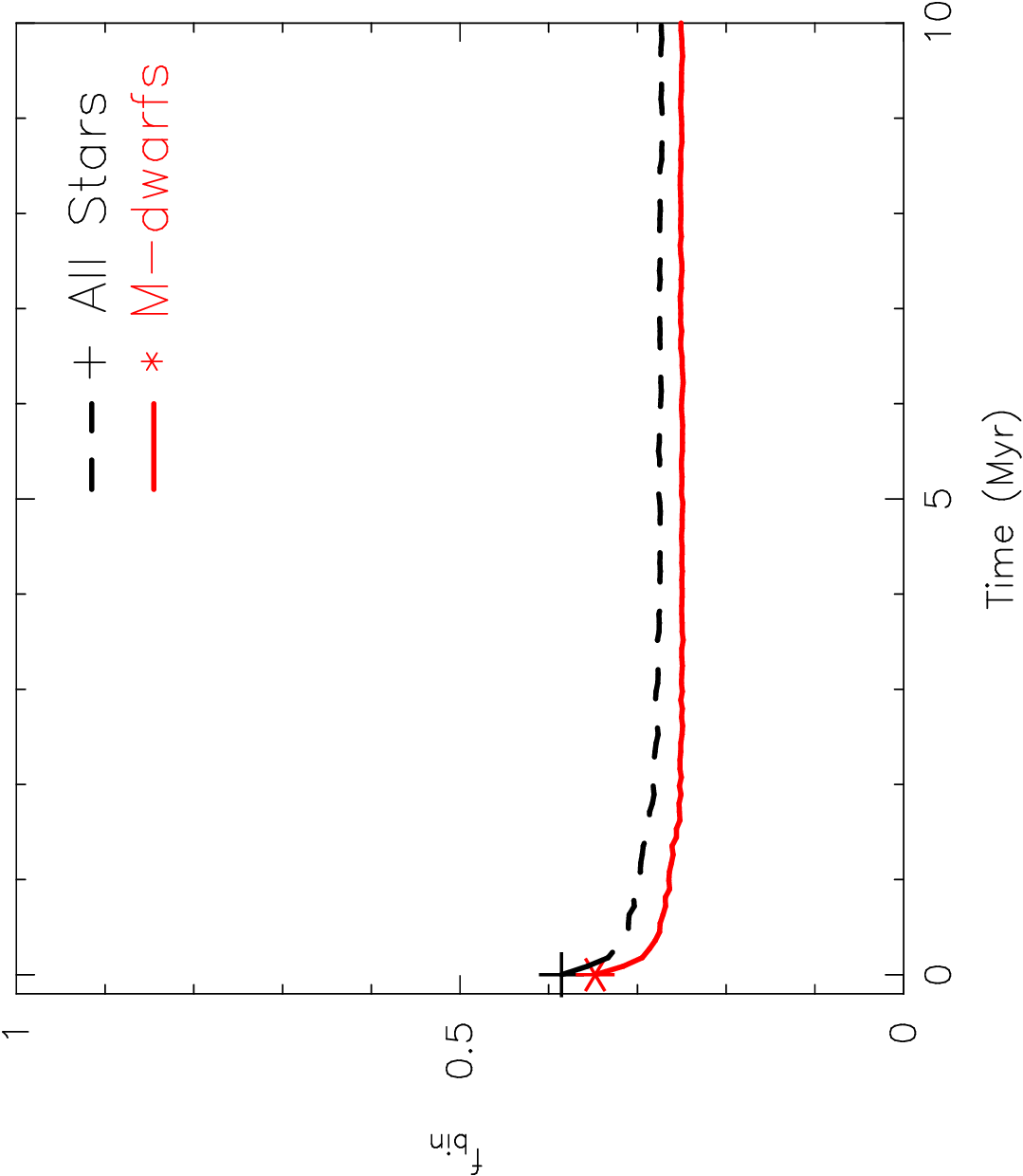}}}
\hspace*{0.6cm} \subfigure[100 per cent binary
    fraction.]{\label{fmult-b}\rotatebox{270}{\includegraphics
    [scale=0.4]{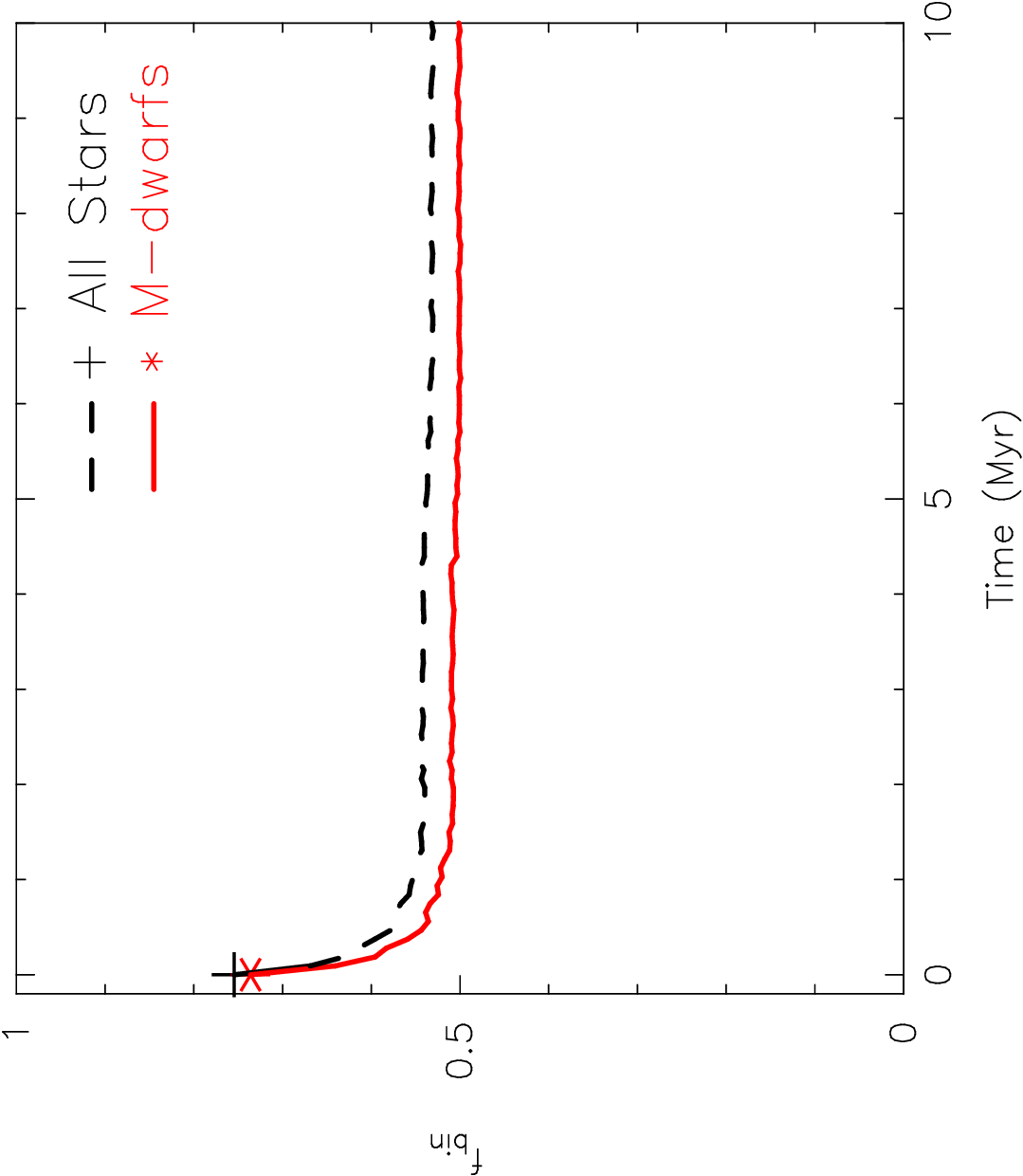}}}
  \end{center}
  \caption[bf]{The evolution of binary fractions with time for a
    $0.1$~pc half-mass radius cluster with (a)  an initial field
    binary fraction; and (b) an initial binary fraction of 100 per
    cent, in which all stars are formed in binary systems. The binary
    fraction for all stars is shown by the dashed (black) line and the
    binary fraction for M-dwarfs is shown by the solid (red) line. The
    initial binary fractions (at cluster birth) are denoted by the
    plus sign  (all stars) and the asterisk (M-dwarfs). In both
    scenarios, a  large proportion of binaries are immediately broken
    up before an equilibrium  state is reached after a few Myrs. The
    final M-dwarf binary fractions after the  cluster has evolved for
    10\,Myrs are 25 per cent for an initial field  binary fraction,
    and 50 per cent for an initial binary fraction of 100 per cent.}
  \label{multiplicity}
\end{figure*}

In Figs.~\ref{fmult-a} and~\ref{fmult-b} we show typical examples of
the evolution of the binary fractions with time for clusters with 
initially field and 100 per cent binary fractions, respectively, with
half-mass radii of $0.1$\,pc.  The two lines represent the evolution of
the binary fraction of {\em all} stars (dashed line), and of M-dwarfs
(solid line).  We define the binary fraction to be the fraction of
multiple (almost always binary) systems compared to the total number
of systems,
\begin{equation}
f_{\rm bin} = \frac{B + T + ...}{S + B + T + ...},
\end{equation}
where $S$, $B$, and $T$ (etc.) are the numbers of single, binary and
triple (etc.) systems, respectively.

% FIGURE ONE -------------------------------------------------------
% ---------------------------------------------------------------------

For both distributions there is a rapid disruption of systems in the
first 0.1~Myr, in which the soft systems are destroyed.  Once the least
bound (i.e. wide) systems are destroyed, the clusters reach an
equilibrium in which the binary fractions remain roughly constant.
This is the same behaviour as found by \citet{Kroupa95a,Kroupa95b}, and
\citet{Kroupa99}.

The destruction of binary systems occurs in roughly a crossing time,
and hence {\em the binary population is set by the densest
configuration of a cluster} before any expansion can occur
\citep{Kroupa00}.   As long as a cluster spends a few {\em initial}
crossing times in a dense configuration, that is enough to process the
binary population.  The current density of a cluster (as long as it is
lower) will merely maintain the population created in the dense state.

What is also clear in Fig.~\ref{multiplicity} is that there is a
mass-dependence in the destruction of systems.  Low-mass binary
systems are more susceptible to destruction as they have a lower
binding energy. The binary fraction  of M-dwarfs is $\sim$~5 per cent
lower than the binary fraction of the other,  more massive stars,
despite the scenario plotted in Fig.~\ref{fmult-b} in which the
various spectral types have the same initial binary fraction.

The wide field-like log-normal
distribution used to generate the initial binary population is not
recovered even before any dynamical evolution.  This is first apparent
in Fig.~\ref{fmult-b},  in which we only obtain a binary fraction of
$\sim$~75~per~cent rather than 100 per cent at time zero, before any
evolution takes place.

It is important to note that the difference is not due to a flaw in
our binary finding algorithm.  Rather, the average distance between
stars in the cluster is around 2000~AU within the half-mass radius.
Therefore binaries with separations of this order or larger -- whilst
generated by the initial conditions -- are not physically associated or
bound within a cluster, and therefore are not  identified as binaries.

\subsubsection{An initially field-like binary fraction}
\label{field_bin_sub}

\begin{figure*}
\begin{center}
\setlength{\subfigcapskip}{10pt}\subfigure[0.1\,pc half-mass radius.]
{\label{field_evol_p1}\rotatebox{270}{\includegraphics[scale=0.4]{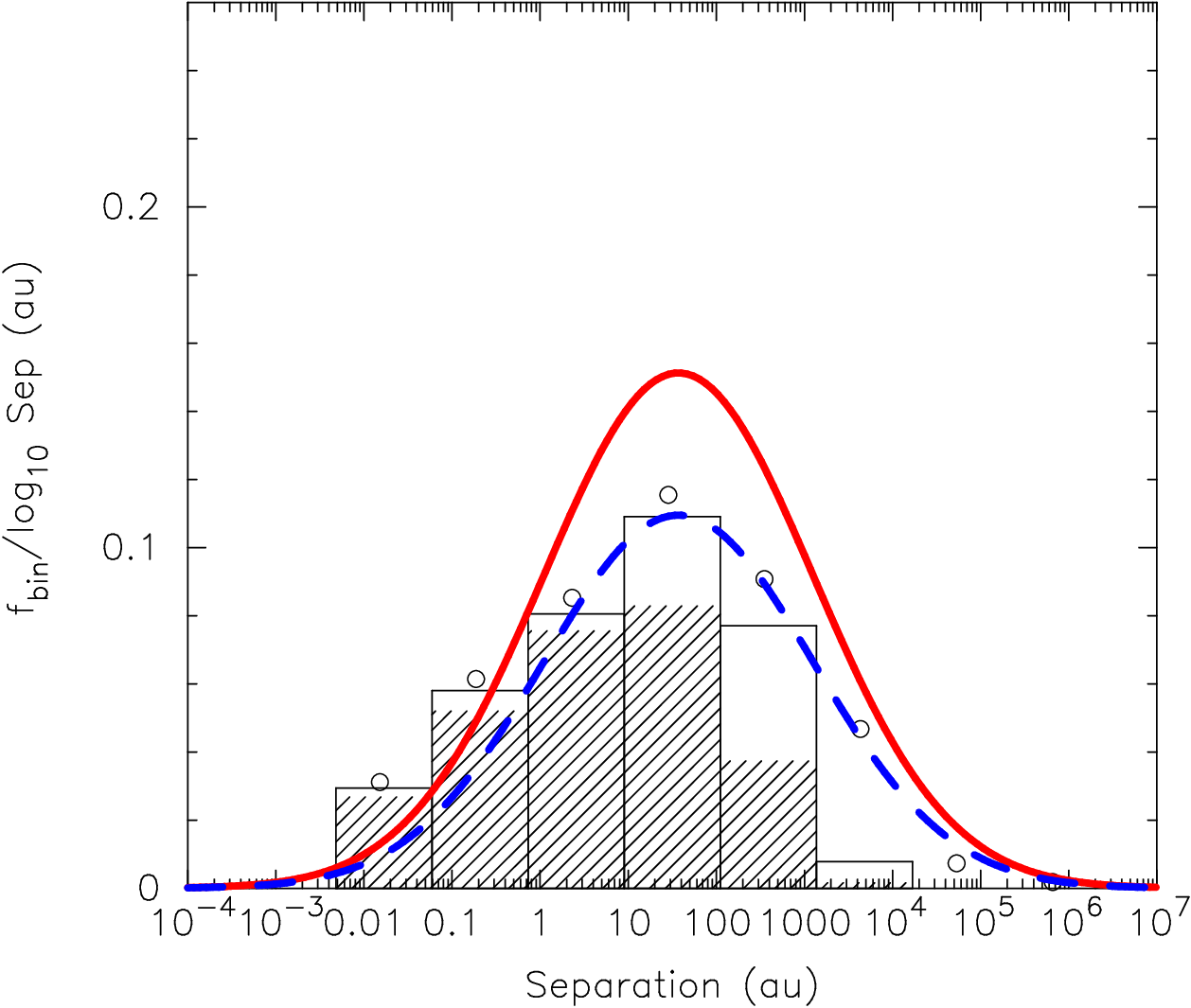}}}
\hspace*{0.3cm} \subfigure[0.2\,pc half-mass radius.]
{\label{field_evol_p2}\rotatebox{270}{\includegraphics[scale=0.4]{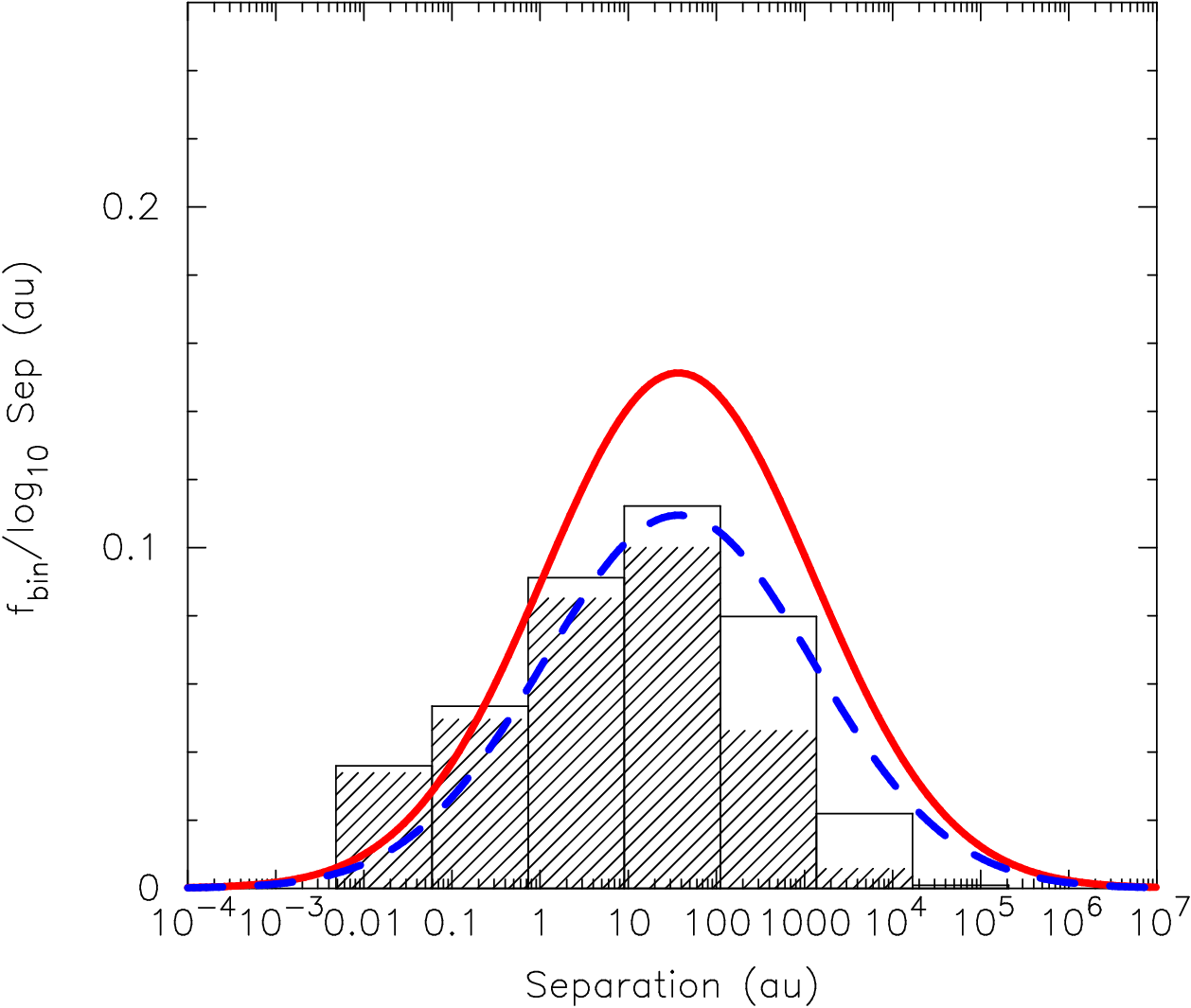}}}
\end{center}
\caption[bf]{The evolution of the separation distribution for a
  cluster with an initially field binary fraction and half-mass radius
  of (a) $0.1$~pc; and (b) $0.2$~pc.  The separation distribution observed for
  field G-dwarfs by \citet{Duquennoy91}  is shown by the solid (red)
  log-normal; the distribution observed for field M-dwarfs by
  \citet{Fischer92} is shown by the dashed (blue) log-normal. The open
  circles in (a) show the initial
  distribution generated by our initial conditions generator, whereas
  in both panels the open histograms show the initial
  binaries and the hatched histogram shows the binaries remaining
  after 1\,Myr, as found by our  algorithm. }
\label{field_evol}
\end{figure*}

In Fig.~\ref{field_evol} we show the binary fraction-separation
distribution with initially field properties in a cluster with
half-mass radii of $0.1$ and $0.2$~pc.  The open histograms show the
distribution at time zero -- before any dynamical evolution of the
systems, the hashed histograms those at  1~Myr 
\citep[roughly the age of Orion;][]{Jeffries07a,Jeffries07b}.  The 
solid (red) lines are the G-dwarf \citep{Duquennoy91} log-normal 
period distribution, and dashed (blue) lines are the M-dwarf 
\citep{Fischer92} log-normal distribution, which are the functions 
used to generate the initial binary populations.  The open circles 
in Fig.~\ref{field_evol_p1} show the binary fractions generated by 
the initial conditions (there is a deviation from the generating 
function at low separations due to the effect of eigenevolution, 
however, as we shall see later this is unimportant as these 
binaries are hard).  The open histograms show the distribution 
of binary fractions found by our binary finding algorithm, which 
are clearly different.

In Fig.~\ref{field_evol} we also compare the separation distributions
of binaries generated with a field binary fraction found by our binary
finder at time zero (open histogram), and (hatched histogram) at 1~Myr. 
Quite clearly there has been significant dynamical destruction of binaries
with separations of $\sim100$~--~$1000$~AU, whilst binaries with
separations $<50$~--~$100$~AU are almost unchanged.  This is the
Heggie-Hills law \citep{Heggie75,Hills75a,Hills75b} in action:  the
hard-soft boundary in our clusters is at a few hundred~AU (this is
also seen by e.g.  \citealt{Kroupa95a,Kroupa95b,Kroupa99} and
described in detail by \citealt{Kroupa08}).

Even the binary fraction of very hard
systems is reduced by the destruction of wide binaries.  There is {\em
very} little evolution in the number of systems with separations below
$1$~AU; however the binary fraction of those systems decreases due to
the increase in the {\em total} number of systems due to the
destruction of wider binaries.  For example, in a cluster with 100
binary systems, 20 of these may be very hard.  The initial binary
fraction of these hard systems would be 20/100~=~20\%.  However, after
the destruction of 20 wider systems, each wider system becomes 40
single stars, and so the binary fraction of very hard systems would be
20/120~=~17\%, despite none of them having been destroyed.

There is very little dynamical processing of initially
  hard systems.  In each cluster a few ($0$ -- $5$) systems with
  initial separations $\ll\,50$~AU are significantly altered or
  destroyed, but most systems retain virtually unchanged separations
  from formation.

We conclude, in common with other authors, that {\em binaries in
dense clusters cannot form with the field separation distribution as
many  field binaries are too wide to have formed in a dense
environment.}  A cluster in which the average separation between
systems is a few thousand AU cannot possibly form systems with
separations greater than this.  Indeed, \citet*{Scally99} find only
$3$ possible binaries in Orion with separations of $1000$ -- $5000$~AU
(they also note that the origin of wide binaries cannot be in Orion-like
clusters).

% FIGURE TWO ---------------------------------------------------------
% --------------------------------------------------------------------

% FIGURE THREE ------------------------------------------------------
\begin{figure*}
%  \begin{center}
\setlength{\subfigcapskip}{10pt} \subfigure[0.1\,pc half-mass
    radius.]{\label{100_dist-a}\rotatebox{270}{\includegraphics
    [scale=0.4]{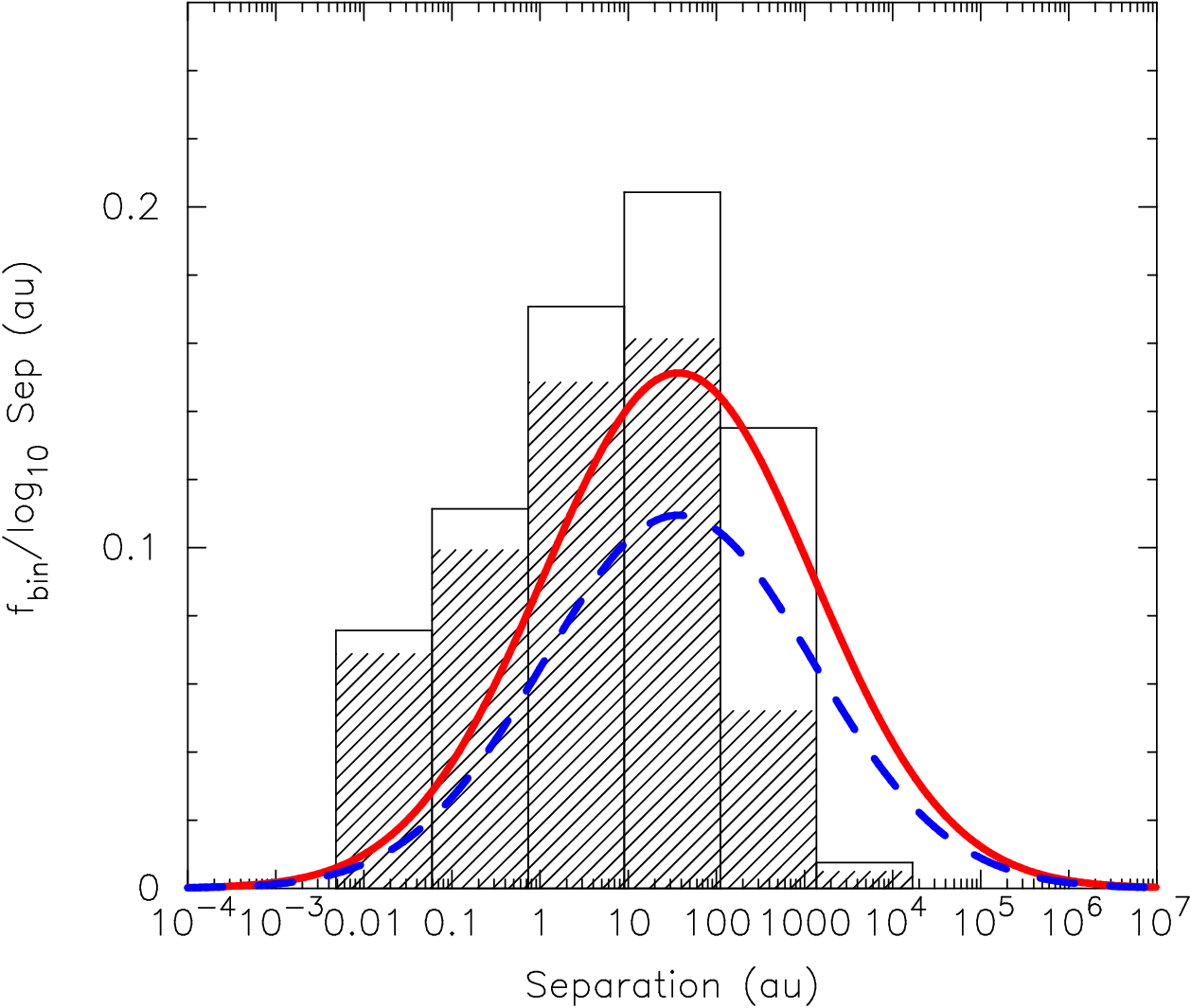}}}
%\hspace*{0.2cm}
    \subfigure[0.2\,pc half-mass
radius.]{\label{100_dist-b}\rotatebox{270}{\includegraphics
[scale=0.4]{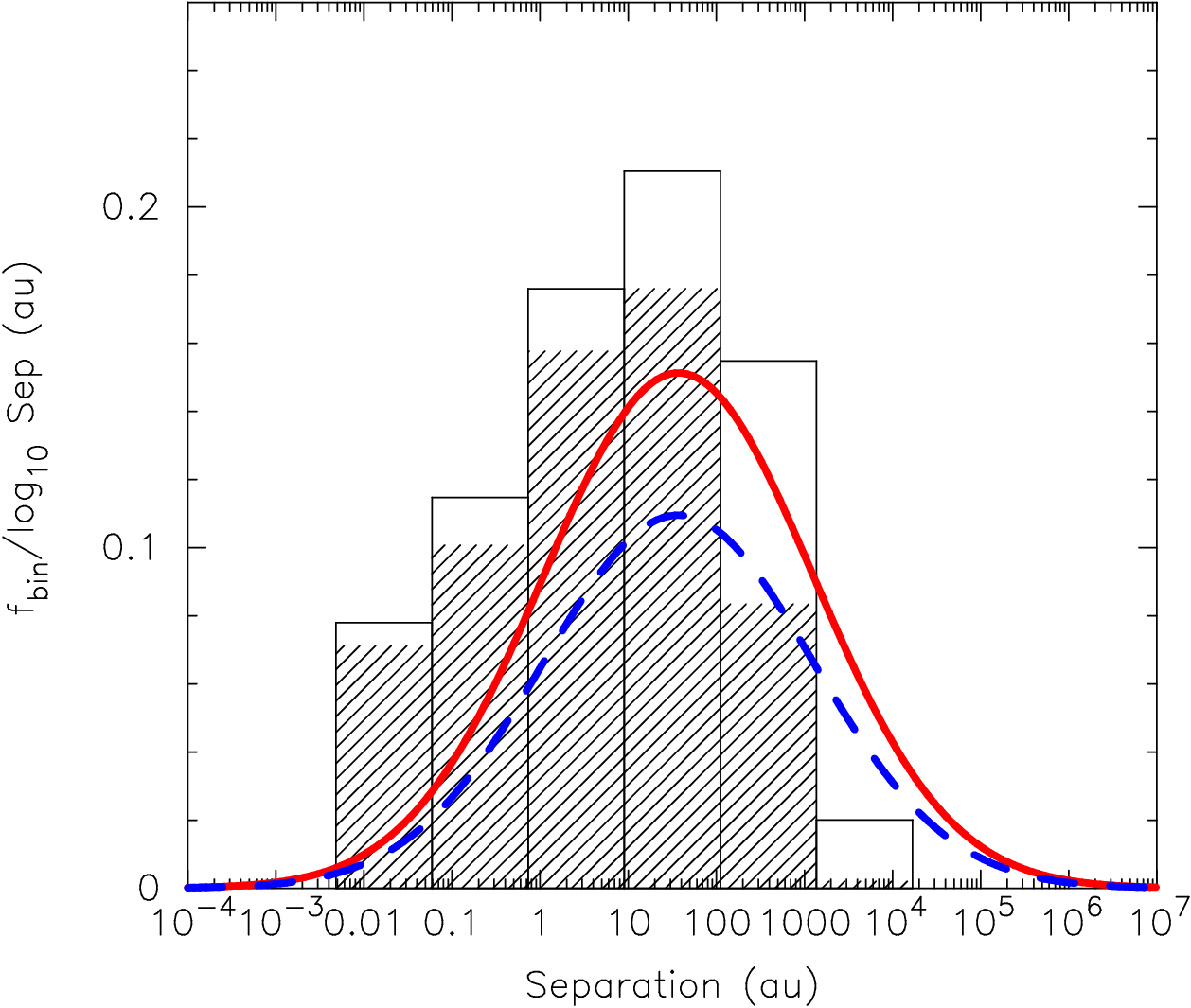}}}
%  \end{center}
  \caption[bf]{The evolution of the separation distributions for
clusters containing $\sim$~1500 stars created with a 100 per cent
binary fraction. The open histograms are the initial distribution and
the hatched histograms  are the distributions after 1\,Myr. We show
the separation distributions for such clusters with initial half-mass
radii of (a)~0.1~pc; and (b) 0.2~pc. The log-normal  fits obtained from
observations of field G-dwarfs \citep[][the solid (red)
line]{Duquennoy91} and field M-dwarfs  \citep[][the dashed (blue)
line]{Fischer92} are also plotted.}
  \label{100_sep_dists}
\end{figure*}
% -----------------------------------------------------------------------

\subsubsection{An initially 100 per cent binary fraction}

Fig.~\ref{fmult-b} appears to show that if the initial binary
fraction is unity in dense clusters, then the effect of dynamical
evolution is to lower the binary fraction to close to the field values
(actually slightly too high for M-dwarfs).  This might suggest that in
dense clusters stars form with a field-like separation distribution,
but with a higher binary fraction (e.g. unity).

\begin{figure*}
%  \begin{center}
\setlength{\subfigcapskip}{10pt}  \subfigure[0.8\,pc half-mass
radius, 1500 stars, field binary
fraction.]{\label{lowdense_dist-a}\rotatebox{270}{\includegraphics[scale=0.4]{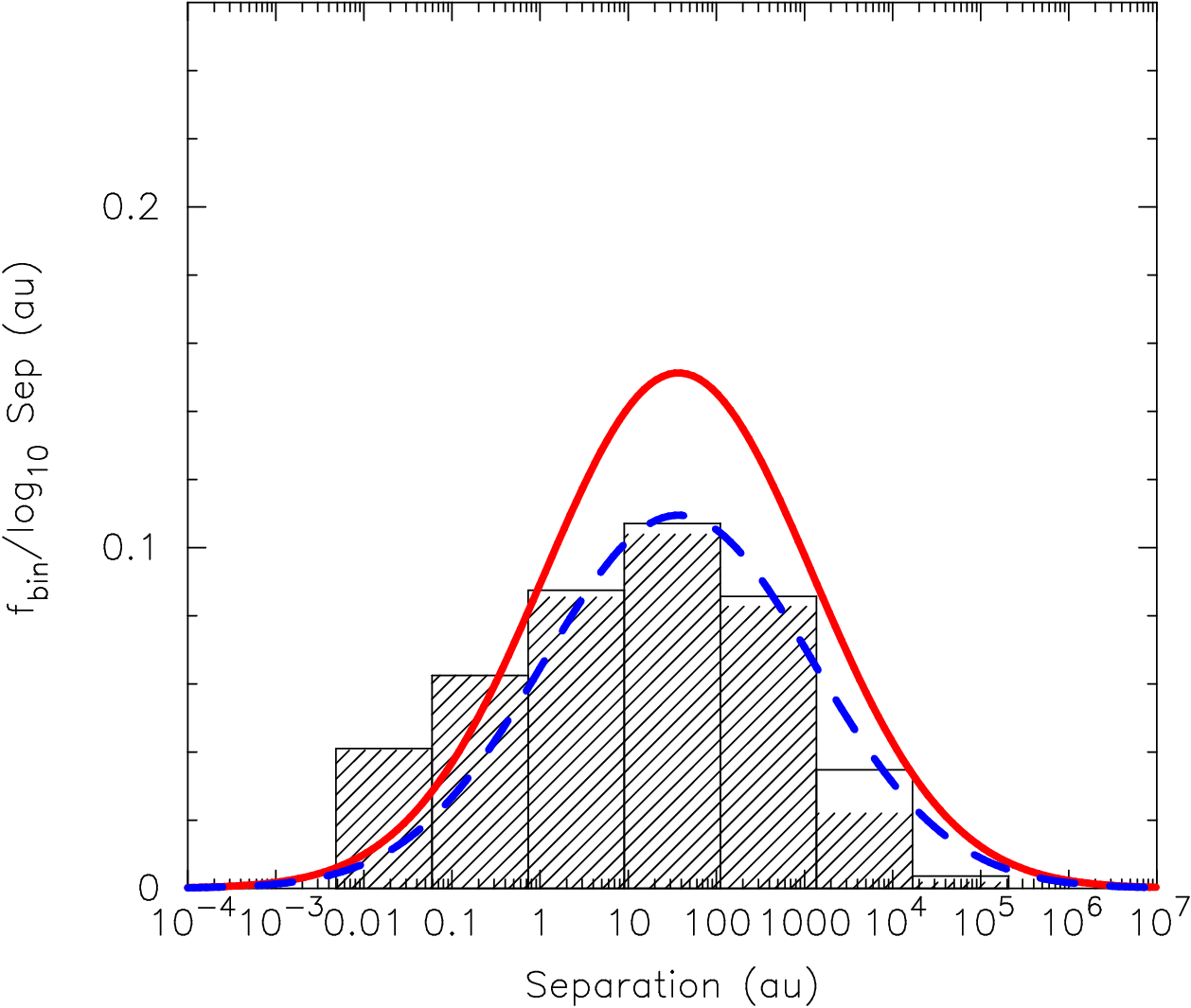}}}
%\hspace*{0.2cm}
    \subfigure[0.8\,pc half-mass radius, 1500 stars, 100 per cent
    binary
    fraction.]{\label{lowdense_dist-b}\rotatebox{270}{\includegraphics[scale=0.4]{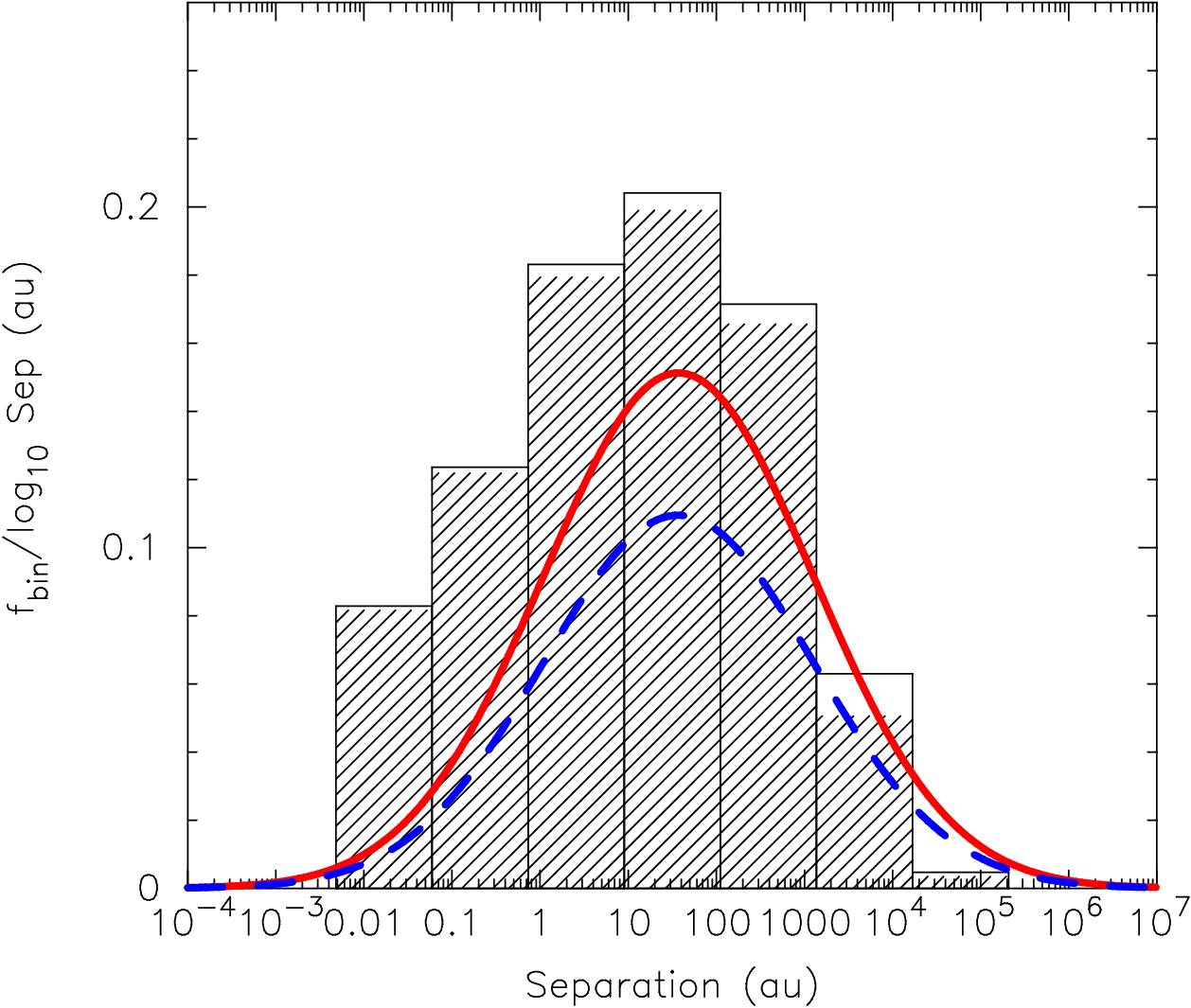}}}
\vspace*{0.5cm}  \subfigure[0.4\,pc half-mass radius, 100 stars,
field binary
fraction.]{\label{lowdense_dist-c}\rotatebox{270}{\includegraphics[scale=0.4]{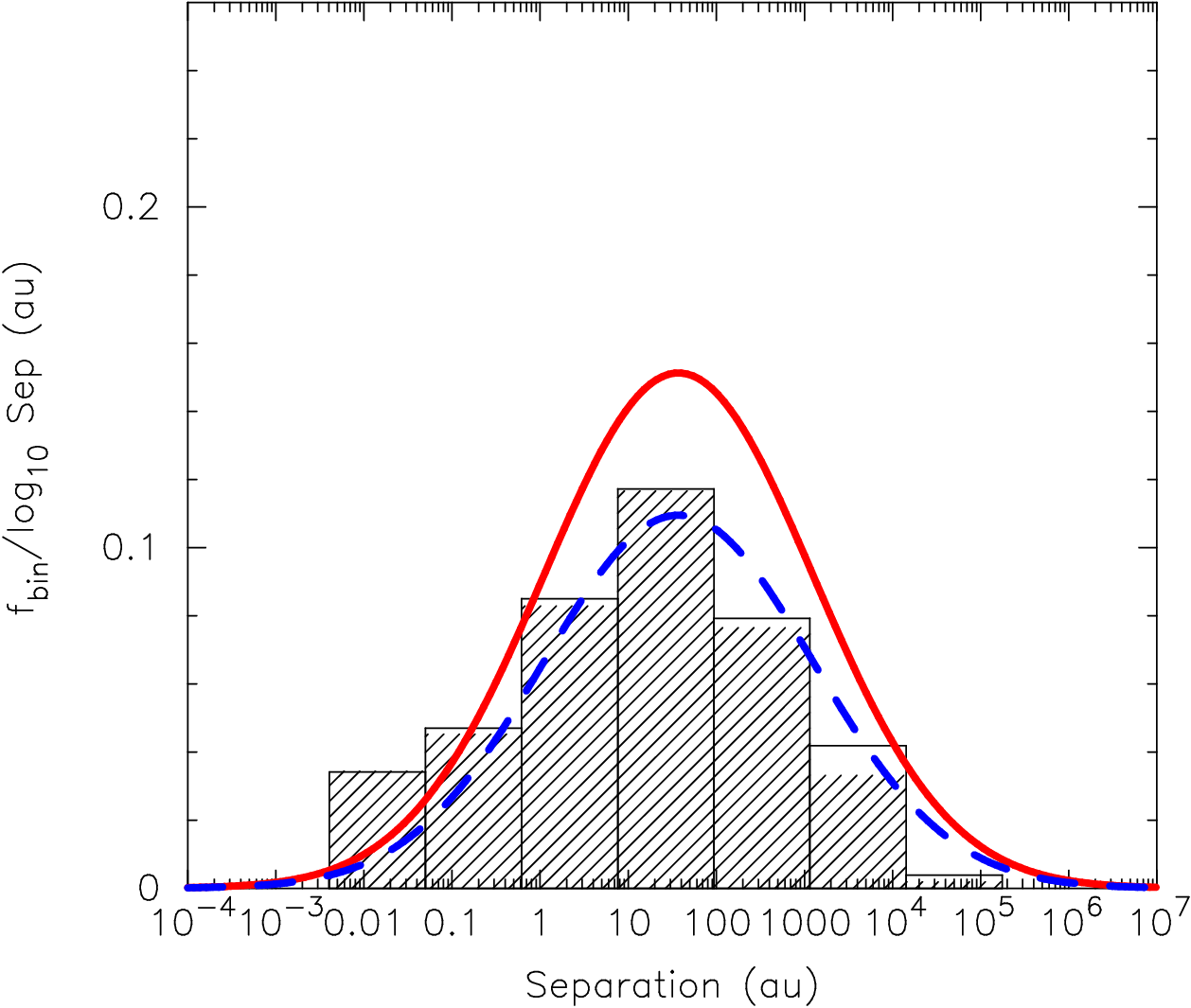}}}
\subfigure[0.4\,pc half-mass radius, 100 stars, 100 per cent
binary
fraction.]{\label{lowdense_dist-d}\rotatebox{270}{\includegraphics[scale=0.4]{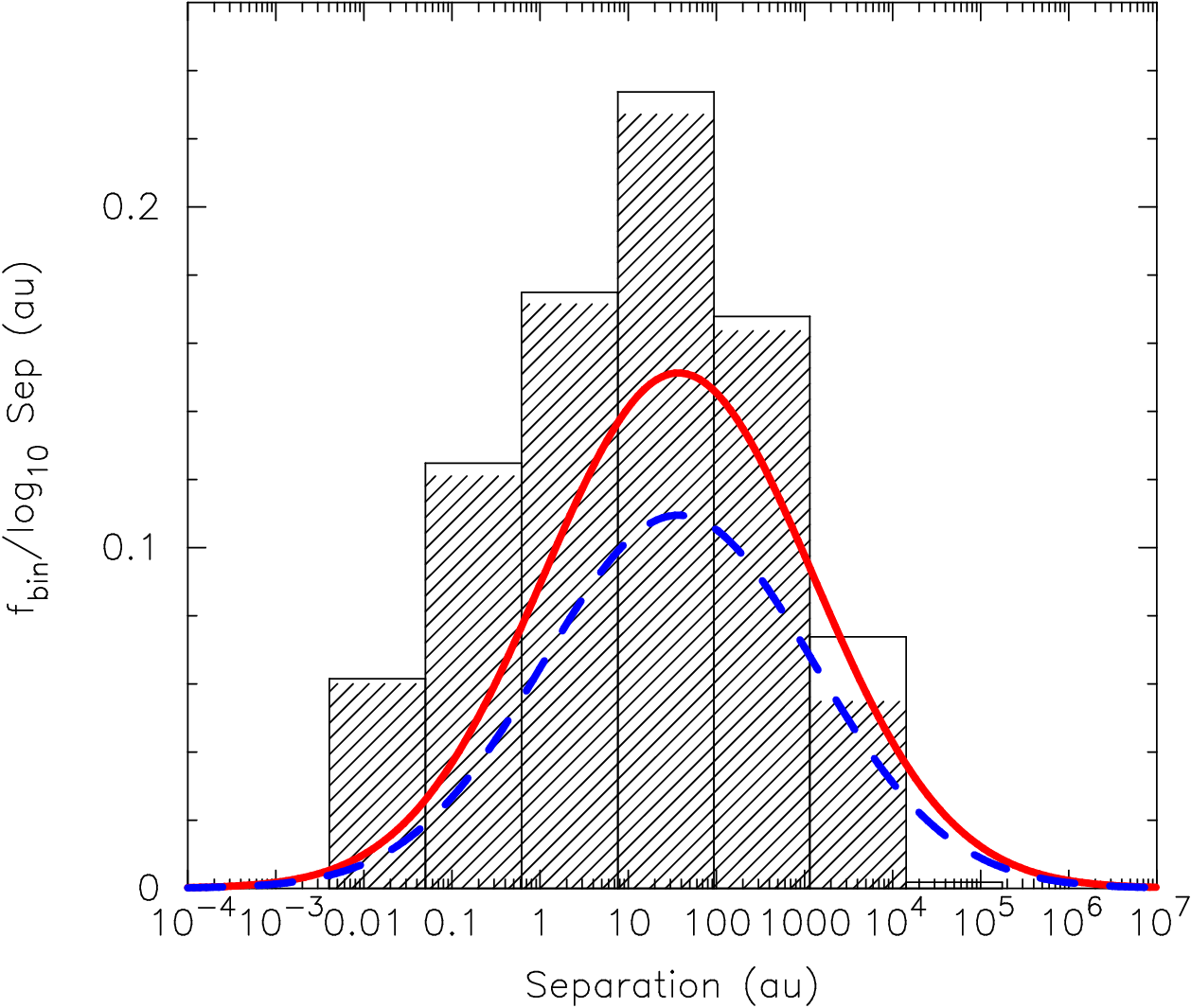}}}
%  \end{center}
  \caption[bf]{ The evolution of the separation distributions for
  clusters with initial half-mass radii of 0.8\,pc, 1500 stars, with
  field-like  and 100 per cent binary fractions ((a) and (b)
  respectively);  and clusters of initial half-mass radii of 0.4\,pc
  containing $\sim$ 100 stars  created with field-like and 100 per
  cent binary fractions  ((c) and (d) respectively). The open
  histograms are the initial distribution and the hatched histograms
  are the distributions after 1\,Myr. The log-normal fits obtained
  from observations of field G-dwarfs \citep[][the solid (red)
  line]{Duquennoy91} and field M-dwarfs \citep[][the dashed (blue)
  line]{Fischer92} are also plotted.}
  \label{lowdense_sep_dists}
\end{figure*}

However, as we show in Fig.~\ref{100_sep_dists} (c.f.
Fig.~\ref{field_evol}), exactly the same  effects occur with a binary
fraction of unity as with a field binary fraction.  Firstly, many of
our generated binaries are unphysically wide, given the cluster's size,
and are not identified as binaries even before the start of the
simulations.  Secondly, the hard-soft boundary is in exactly the same
place and so many binaries with separations  $>50$ to a few hundred~AU
are dynamically disrupted.

Our initial conditions also produce too many low-separation
binaries which are unaffected by dynamical evolution.  This suggests
that the initial {\em hard} binary population should look similar to
the field (at least after mixing with clusters of lower densities and
field stars -- see Section~\ref{discussion}).

It should be noted that high-density clusters -- even though they
start in virial equilibrium -- expand significantly due to the input
of energy into the cluster from the disruption of significant numbers
of binary systems.  

%xxxxxxxxxxxxxxxxxxxxxxxxxxxxxxxxxxxxxxxxxxxxxxxxx

\subsection{Low-density clusters}

We now examine the dynamical evolution of binary systems in
low-density environments.  In Fig.~\ref{lowdense_sep_dists} we show
the evolution of an initially field-like initial binary population in
a cluster with $\sim 1500$ members with a half-mass radius of
$0.8$~pc, and of a cluster with $\sim 100$ members and a half-mass
radius of $0.4$~pc.  These clusters both have densities of 
$\sim~200$~--~$300$\,M$_\odot$~pc$^{-3}$ (compare with the clusters illustrated in
Fig.~\ref{field_evol} which have densities of $\sim 10^4$\,M$_\odot$
pc$^{-3}$). We show the binary separation  distributions initially
(zero time, rather than the initial conditions), and at 1~Myr  for the
clusters.

What is clear in all panels in Fig.~\ref{lowdense_sep_dists} is
that most very wide binaries ($> 10^4$~AU) never survive (even if they form) 
even in low-density clusters.  The hard-soft boundary has clearly shifted in
the low-density clusters compared to the high-density clusters
considered above, but not enough to allow these extremely wide
binaries to survive.

From Figs.~\ref{field_evol} and~\ref{100_sep_dists} (and inspection of the data) we
estimate the hard-soft boundary in a dense $N \sim 1500$, $r_{1/2}~=~0.1$~pc 
cluster to be $\sim 50$ -- $100$~AU.  For the low density
cluster with $N \sim 1500$, $r_{1/2} = 0.8$~pc, Eqn.~\ref{eqn:hs} would
suggest that the hard-soft boundary should be at around $500$~AU.
However, inspection of Fig.~\ref{lowdense_sep_dists} seems to show
that the limit of the destruction of intermediate binaries is around
$5000$~AU, about ten times larger than expected.

The reason for this discrepancy is that the low density cluster is
still dynamically young at the 1~Myr ages illustrated in the previous
figures.  The crossing times of the dense clusters are only 10s~kyr
meaning that the clusters are many crossing times old.  However, the
crossing times of the low density clusters are around $0.4$~Myr,
meaning that they are dynamically very young. As illustrated in 
Fig.~\ref{multiplicity} and described above, it takes a few crossing
times to reach an equilibrium in which all of the soft systems have
been destroyed and the low density clusters have not had long enough
to do this by 1~Myr.  This also means that the low density clusters
are not dynamically mixed and the binary fraction has a radial dependence 
(we will describe this in detail in the next section).

%==========================================================================
\section{The initial conditions of the Orion Nebula Cluster}
\label{ONCoriginal}

The Orion Nebula Cluster (ONC) is observed to have a
half-mass radius of $\sim 0.8$~pc and around 2000 -- 3000 members 
\citep{McCaughrean94,Hillenbrand98,Kohler06}.  This makes the ONC a
low-density cluster following our definitions above.  However, this is
the {\em current} state of the ONC.  As shown by \citet{Bastian08}
many clusters seem to undergo an early dense phase and then rapidly
expand \citep[possibly due to gas expulsion, see][]{Goodwin06}.
In this section we examine the binary properties of the ONC and its
dynamical state to attempt to infer the initial conditions of the ONC.

\citet{Reipurth07} carried out a survey of intermediate/wide (visual) 
binaries in the ONC. They looked at the separation distribution in the 
range of $\sim$~68\,AU to 676\,AU.  As noted in Section~\ref{field_bin_sub}, 
\citet{Scally99} find only $3$ possible binaries in Orion with separations 
greater than $1000$~AU.  Therefore, the \citet{Reipurth07} sample covers most of
the range of intermediate to wide binaries in the ONC (presumably
there are also a small number of binaries in the range 700 -- 1000~AU).

The immediate and obvious conclusion to draw from this is
that ONC-like clusters -- even at low-density -- cannot be the source
of a significant number of binaries with separations $> 1000$~AU in the field.

\begin{figure*}
  \begin{center}
\setlength{\subfigcapskip}{10pt} \subfigure[0.1\,pc half-mass radius,
    100 per cent binary
    fraction.]{\label{reipurth-a}\rotatebox{270}{\includegraphics
    [scale=0.4]{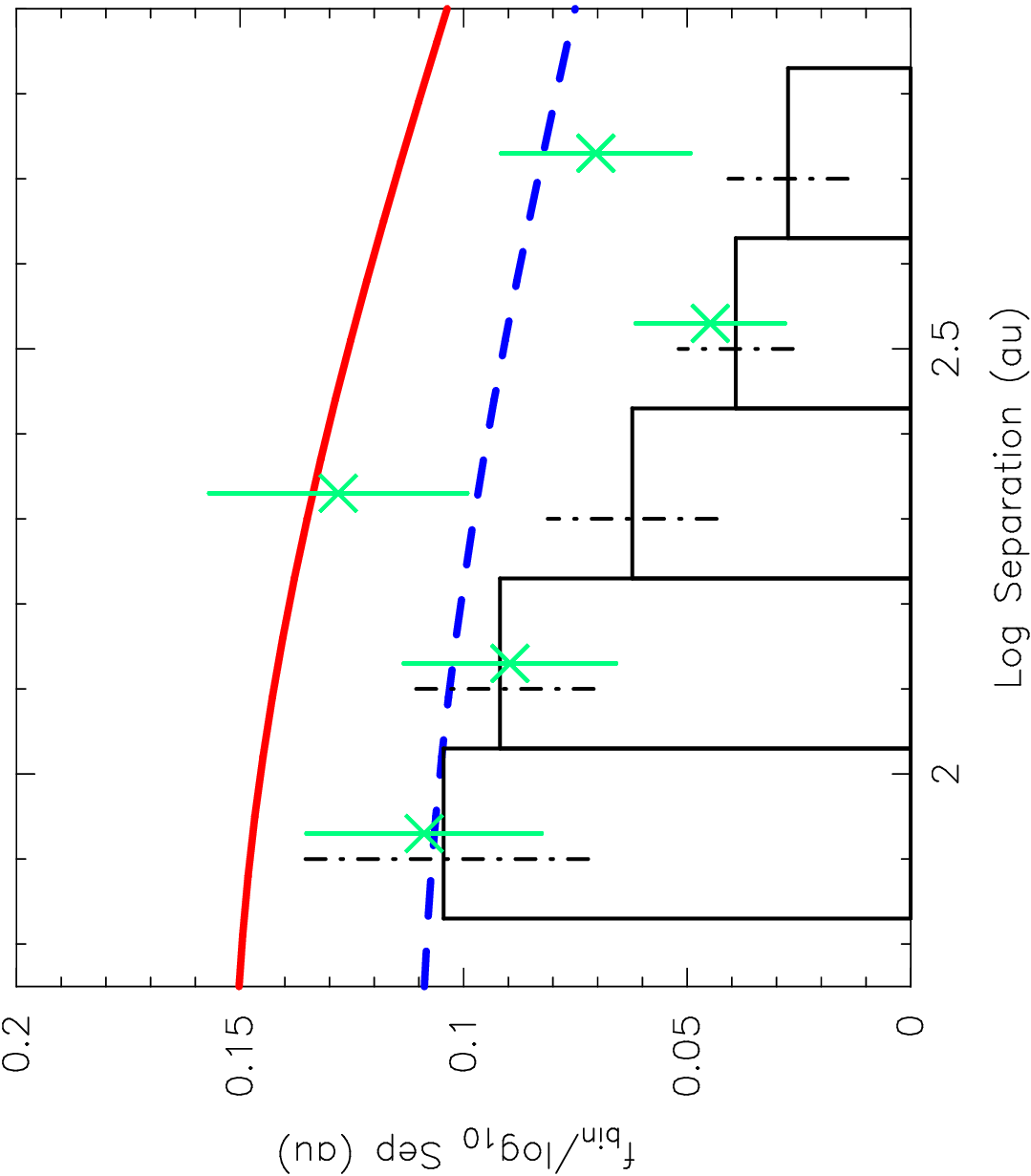}}}
\hspace*{0.6cm} \subfigure[0.2\,pc half-mass radius, 100 per cent
    binary
    fraction.]{\label{reipurth-b}\rotatebox{270}{\includegraphics
    [scale=0.4]{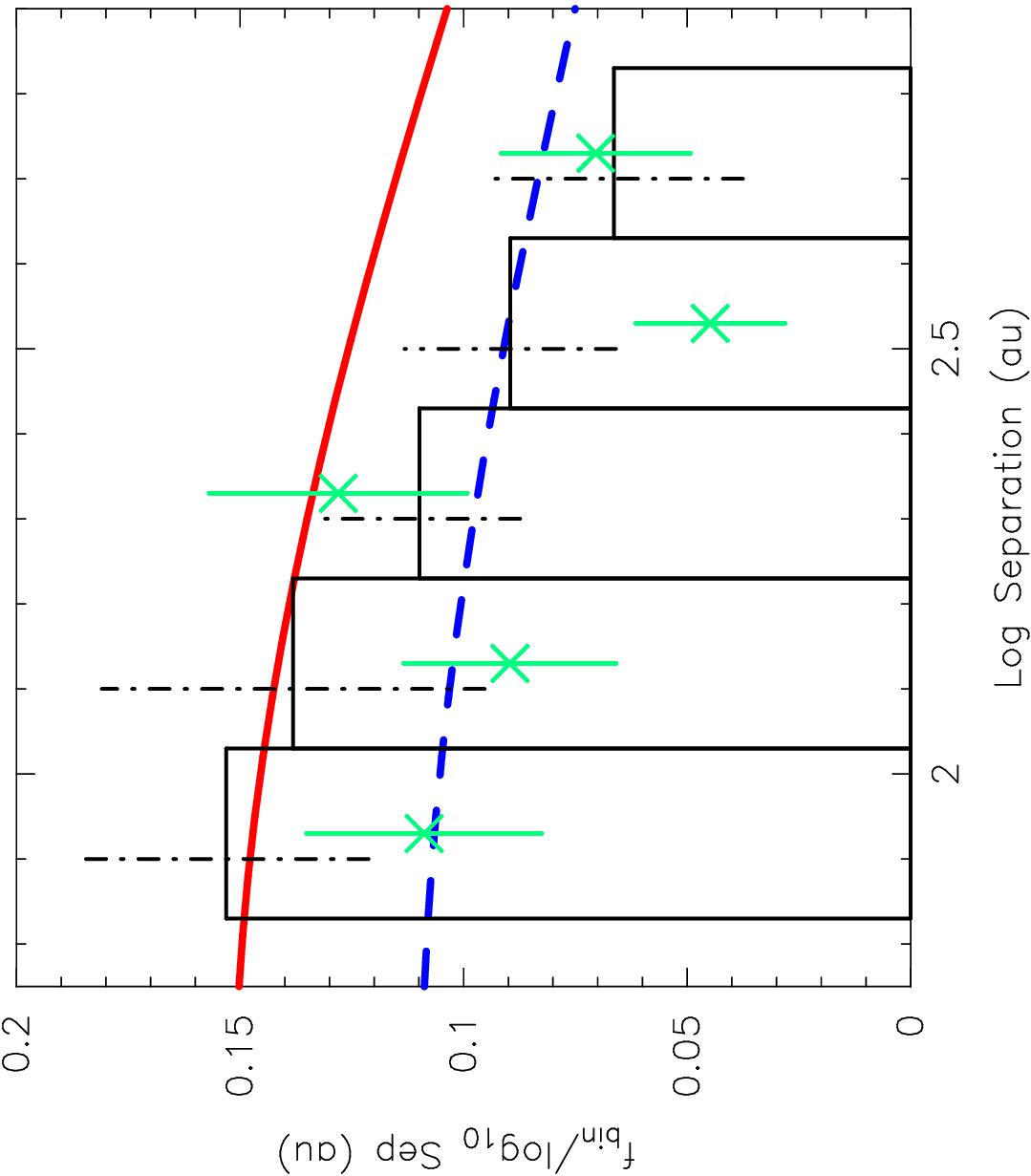}}}
\vspace*{0.5cm} \subfigure[0.4\,pc half-mass radius, field binary
    fraction.]{\label{reipurth-c}\rotatebox{270}{\includegraphics
    [scale=0.4]{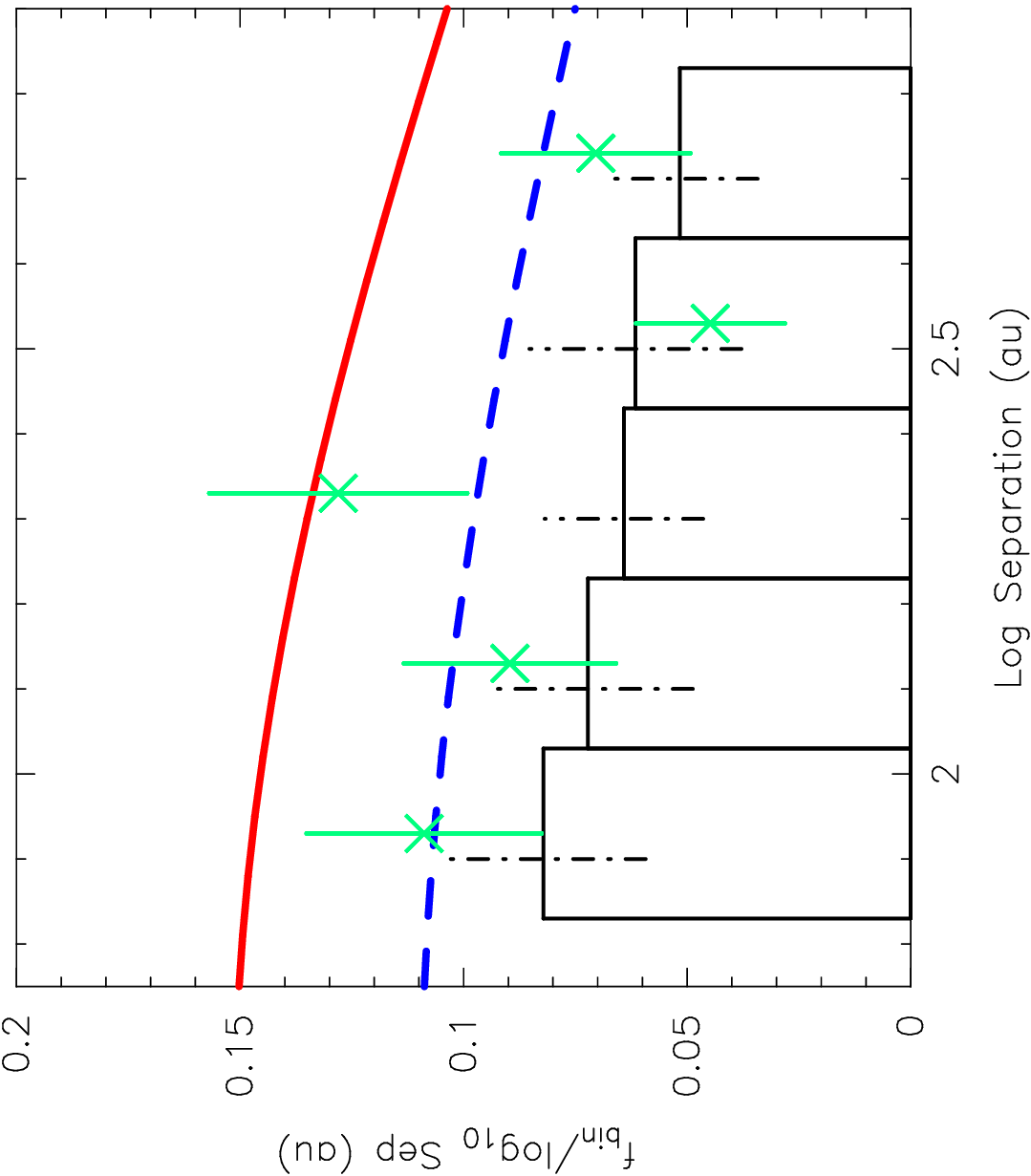}}}
\hspace*{0.6cm} \subfigure[0.8\,pc half-mass radius, field binary
    fraction.]{\label{reipurth-d}\rotatebox{270}{\includegraphics
    [scale=0.4]{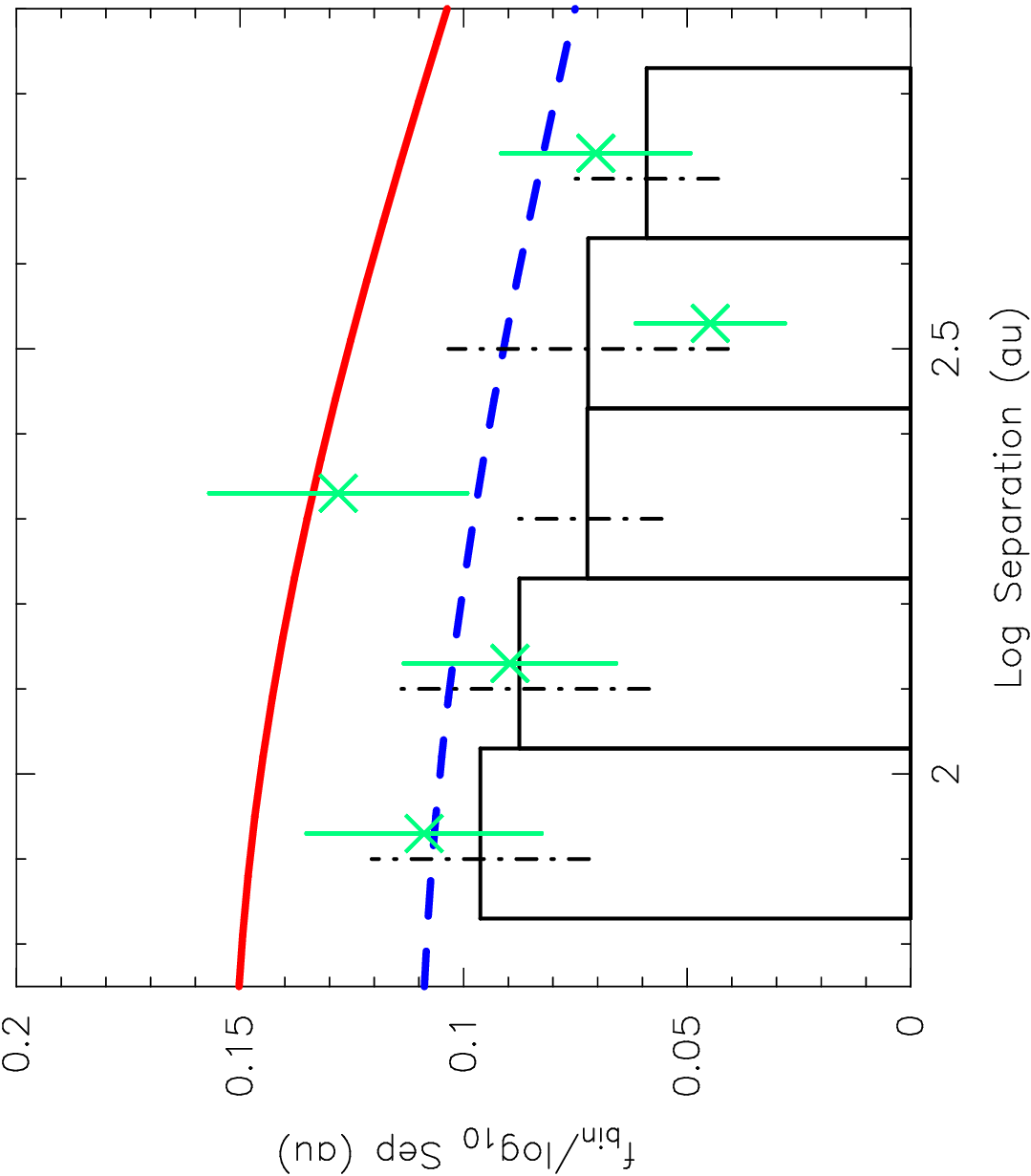}}}
  \end{center}
  \caption[bf]{Comparison with fig.~7 from \citet{Reipurth07}. Their
data are shown by the  (green) crosses, together with their
uncertainties. Data from our simulations for a cluster age of 1\,Myr
and cluster membership of $\sim$\,1500 stars (similar to the ONC) 
are shown by  the open histograms (uncertainties shown by the
dash--dot lines and offset from the centre  of each bin for
clarity). As in our Figs.~\ref{field_evol},~\ref{100_sep_dists}~and~\ref{lowdense_sep_dists},
the log-normal distributions for field G-dwarfs \citep[][the solid
(red) line]{Duquennoy91},  and field  M-dwarfs \citep[][the dashed
(blue) line]{Fischer92} are plotted. We show the results for three
different initial cluster half-mass radii. Panels (a) and (b) show the
results for an  initial binary fraction of 100 per cent and half mass
radii of 0.1\,pc and 0.2\,pc respectively.  Panels (c) and (d)
show the results for an initial field binary fraction and a half mass
radii of 0.4\,pc, and 0.8\,pc respectively.}
  \label{reipurth}
\end{figure*}

The \citet{Reipurth07} data are shown in  Fig.~\ref{reipurth}; they
place their data into bins of width 0.2\,log AU and the numbers of 
binaries in each bin are shown by (green) crosses.  We also plot the
log-normal  distributions  for G-dwarfs and M-dwarfs.  Note that most
of the stars observed by \citet{Reipurth07} are M-dwarfs, and so it is
the (lower) M-dwarf distribution that should be compared to the data
rather than the G-dwarf distribution.

We over-plot the average of our ensemble of simulations for clusters
with 100 per cent initial binary fractions and half-mass radii of
$0.1$ (Fig.~\ref{reipurth-a})  and $0.2$~pc (Fig.~\ref{reipurth-b}),
and an initially field binary fraction and half-mass radii of $0.4$~pc
(Fig.~\ref{reipurth-c}) and $0.8$~pc (Fig.~\ref{reipurth-d}),
shown by the histograms in each panel.  Each histogram shows the
average for that bin of 10 simulations, with the dot-dashed lines
showing the standard deviations over the 10 realisations.

From inspection of  Fig.~\ref{reipurth} it is clear that the
observations of the ONC can be matched by both an initially high binary
fraction in a high-density  cluster, or an initially field-like
distribution in a low-density cluster.  

The current half-mass radius of Orion is $\sim 0.8$~pc, a factor of
$\sim 5$ larger than that inferred from an initially high binary
population.  Despite the fact that a field-like population can match
the observations we strongly prefer the interpretation that Orion
started with a far higher binary fraction in this separation range
which has dynamically evolved into the current distribution.  The
reasons for this are threefold.

$\bullet$ Firstly, the central region of Orion (0 -- 5\,pc) is
thought to be expanding\footnote{On  larger scales ($\sim$20\,pc),
the region around the ONC may be undergoing cold collapse, as observed by
\citet{Feigelson05} and \citet{Tobin09}.},  based on the 1D velocity dispersion of $\sim
2.5$ km\,s$^{-1}$ \citep{Jones88}.   The 3D velocity dispersion is
$\sim 4.3$ km\,s$^{-1}$ \citep*{Kroupa01a,Olczak08}, which is too
large for it to be in virial equilibrium \citep[although we note that
the effect of binaries in a cluster is to increase the observed
velocity  dispersion;][]{Kouwenhoven08} -- a $0.8$~pc half-mass radius
cluster  with a mass of $1500$\,M$_\odot$ has a virialised velocity
dispersion of  $\sim 2.8$ km\,s$^{-1}$
\citep[see][]{Kroupa99,Kroupa01a}.  Clusters are expected to expand
early in their lives due to the effects of gas expulsion
\citep[see][]{Bastian06,Goodwin06}.  Indeed, a factor of $6$ increase
in radius is not unusual for a low effective star formation efficiency
cluster \citep[see][in particular their fig.~2]{Goodwin06}.   The
current velocity dispersion of $4.3$ km\,s$^{-1}$  implies an initial
size of the cluster of $<0.2$~pc\footnote{Whilst an initially
super-virial cluster expands its velocity dispersion does
decrease. However, to have expanded to 0.8~pc with a velocity
dispersion of $4.3$ km\,s$^{-1}$ in 1~Myr the young ONC must have been
significantly super-virial, probably with an effective star formation
efficiency (eSFE) of $\sim 0.3$ \citep{Kroupa01a, Goodwin06}.  Such a
low eSFE would suggest that the velocity dispersion would not decrease
significantly during the expansion as the cluster would be globally
unbound \citep{Goodwin06}.}.  Therefore, its initial configuration
{\em must} have been denser than is currently observed -- and, as  we
have seen, it is the initial, rather than current density that is all
important for processing the initial binary population.
\citet*{Scally05} suggest  from their detailed study of the dynamics
of the Orion cluster that it may well have initially been $10$ --
$100$ times denser in the past; we suggest  that it was around $100$
times denser, in broad agreement with them.

\begin{figure*}
\setlength{\subfigcapskip}{10pt} \subfigure[0.1\,pc half-mass
    radius, 100 per cent binary fraction.]{\label{radmult_p1}\rotatebox{270}{\includegraphics
    [scale=0.4]{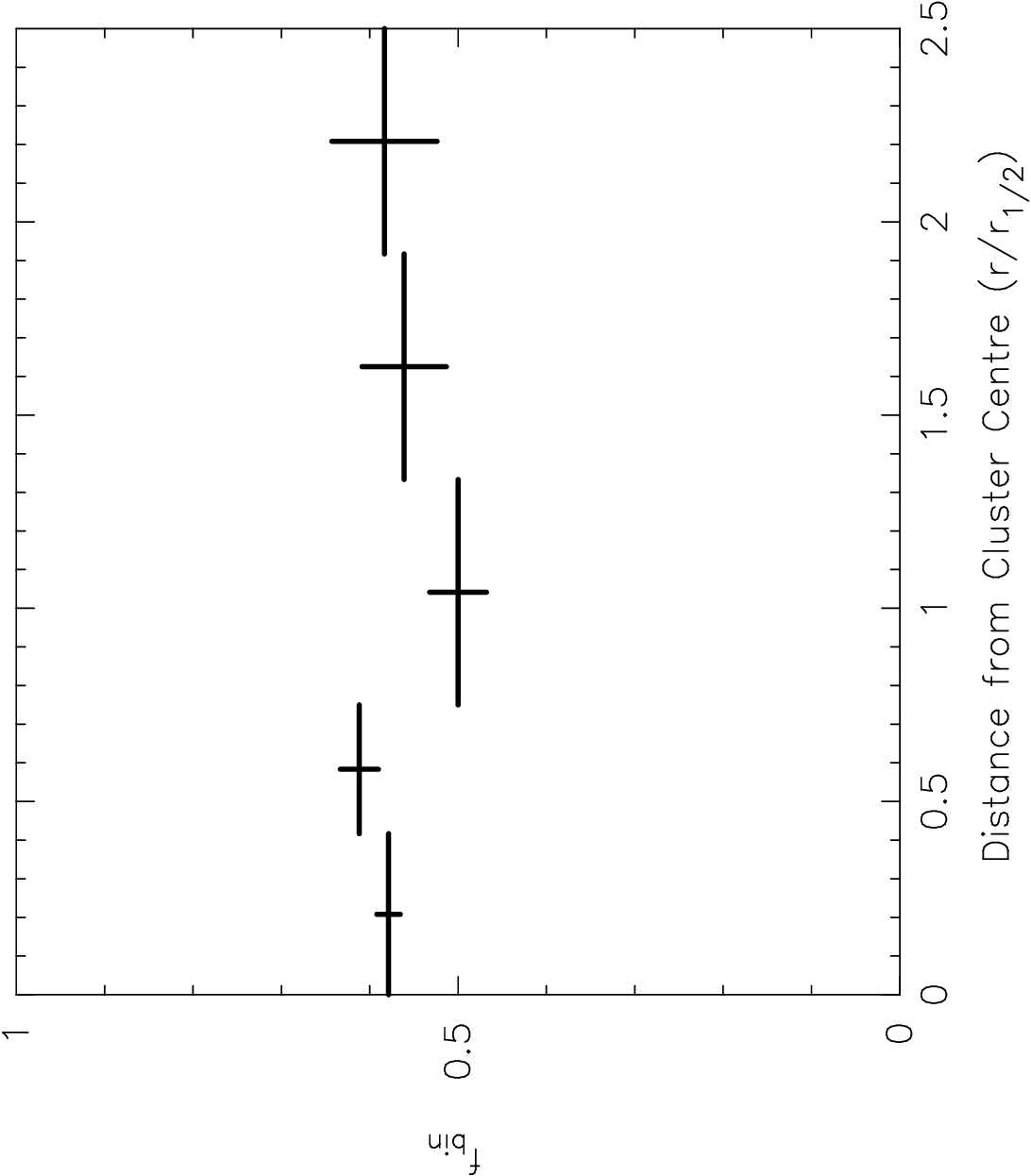}}}
\hspace*{0.2cm} \subfigure[0.8\,pc half-mass
    radius, field binary fraction.]{\label{radmult_p8}\rotatebox{270}{\includegraphics
    [scale=0.4]{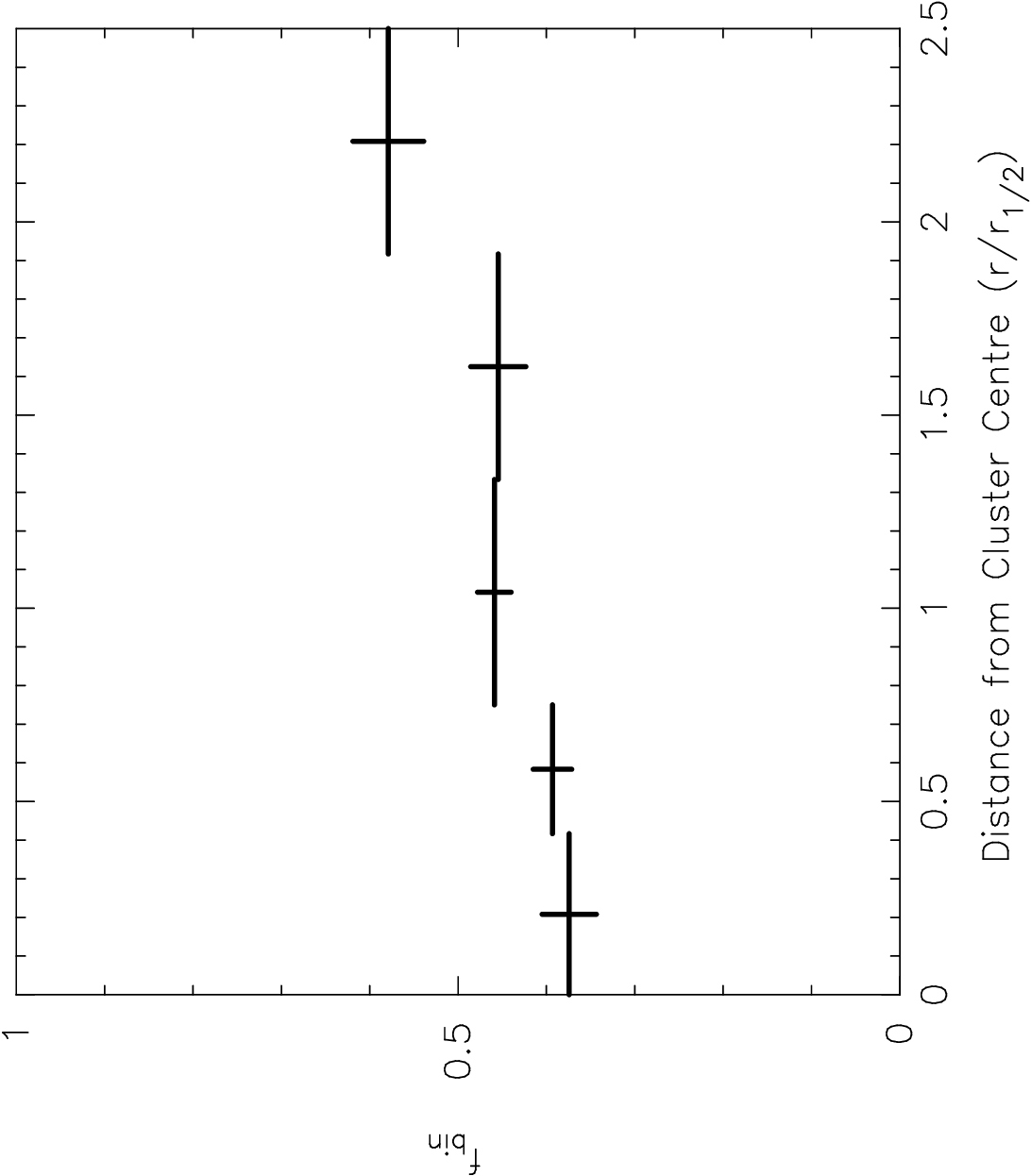}}}
  \caption[bf]{The binary fraction as a function of radius after
  1\,Myr of evolution for clusters containing 1500 stars but with
  different initial binary fractions and different initial densities:
  (a) with an initial half-mass radius of 0.1\,pc and a binary
  fraction of 100 per cent, and (b) with an initial half mass radius
  of 0.8\,pc and a field-like binary fraction. We plot the binary
  fraction of all stars in the cluster against the distance from the
  centre in half-mass radii. This enables a direct comparison with the
  ONC data in figs.~10~and~11 in \citet{Kohler06}.}
  \label{radmult}
\end{figure*}

$\bullet$ Secondly, the binary population observed by \citet{Kohler06} and
\citet{Reipurth07} in Orion appears to be well-mixed. \citet{Kohler06} find 
very little difference in the binary fraction (more correctly, the companion 
star frequency, see their figs.~10 and~11) with radius.  This is difficult to
explain if Orion started large, as the inner regions should have been
more heavily processed than the outer regions as the {\em current}
crossing time for the cluster is of order its age \citep[see
e.g.][]{Kohler04}.   However, if the cluster started out $\sim 100$
times denser than it is now, then it is significantly dynamically
older than its current size would suggest  \citep[as would
be the case for expanding clusters;][]{Bastian08}.  Hence its
binary population would be well-mixed and evolved within a few hundred
thousand years (as discussed above) and no very significant differences
between the binary fractions in the inner and outer regions would be
expected. In Fig.~\ref{radmult} we show the radial variation of
the binary fraction at 1~Myr in a dense ($r_{1/2} = 0.1$~pc) cluster
with an initial binary fraction of 100 per cent and low-density
($r_{1/2} = 0.8$~pc) cluster with an initial field-like binary
fraction.  (Note that the dense cluster expands, so
that after 1\,Myr $r_{1/2} = 0.2$~pc, whereas the sparse cluster
retains its initial half-mass radius.) Our simulations show a trend 
towards an increasing binary fraction with
 radius  for the low-density (dynamically young) cluster,
 and no trend  in the high-density (dynamically old) cluster.

\citet{Reipurth07} do find that the ratio of wide-to-close binaries 
increases with radius within the inner~pc, and is flat beyond this 
(see their fig.~9; close binaries are 66~--~225~AU, and wide 
binaries are 225 -- 670~AU).  However, easily within the errors the 
ratios are flat beyond the inner 0.5~pc, and it is only within the 
inner 0.5~pc that there are significantly fewer wide binaries than close ones.  
This is not at all unexpected.  While the Orion cluster has been expanding 
from its proposed denser initial state, it will still undergo dynamical processing 
as wider binaries are always more susceptible to destruction in the inner regions 
of a cluster than in the outer regions.  If Orion was always at its current size, 
then the wide binary population in the outer regions is very close to the {\em initial} 
population, whilst in the inner regions it has been processed (the situation in 
Fig.~\ref{radmult_p8}).  However, if Orion was originally far denser, then the 
wide binary population has been processed in every region of the cluster, only 
somewhat more so in the inner regions (the situation in Fig.~\ref{radmult_p1}) 
and at no location in the cluster do we see the initial wide binary population.  
This leads to our third point.

$\bullet$ Finally, \citet{Scally99} find very little evidence of binaries
with separations $> 1000$~AU in the ONC.  If the ONC formed at low
density, it must have formed with almost no binaries with separations 
$> 1000$~AU as a low density cluster could not have been able to destroy a population
of binaries this wide in 1~Myr.  But the lack of binaries with separations $> 1000$~AU
can easily be explained if the ONC was far more dense in the past as
it would have destroyed the majority of binaries with these separations. 
In addition, that we do not see binaries with separations $> 1000$~AU even in the very 
outer regions of the cluster suggests that the {\em entire} cluster 
has been dynamically processed.

Whilst none of these arguments show conclusively that the ONC has
  expanded from an initially dense stage, we think they are at least
  an indication.  However, we would also note that \citet[][ApJL, submitted]{Allison09} 
  can explain the formation of the Trapezium if the ONC was
  once $\sim 100$ times denser.

\section{Discussion}
\label{discussion}

Binary fractions and multiplicities are clearly strong constraints on
theories of star formation \citep{Goodwin07}.  However, it is
often very difficult to compare the outcomes of theory with
observations.  In particular, the outcome of star formation theories
cannot be directly compared to the field \citep*[as attempted by
e.g.][]{Goodwin04,Bate09} as (a) there is significant
dynamical evolution in dense clusters which will alter the star
formation products, and (b) the field is the sum of many star forming
regions and modes \citep[e.g.][]{Brandner98}.

We have examined two density regimes -- high-density clusters 
($10^4-10^5$\,M$_\odot$ pc$^{-3}$), and low-density clusters 
(${\cal O}(10^2)$\,M$_\odot$~pc$^{-3}$).  We have argued that these two
groups bracket the vast majority of densities in cluster-forming star
formation.  We have seen that {\em neither} low- or high-density clusters allow
binaries with separations $> 10^4$~AU to exist.  And high-density
clusters do not allow binaries with separations $> 10^3$~AU to exist.  However,
observations of the field show that such binaries apparently do exist
in significant numbers.  This raises a number of questions about the
origin of the field population.

\subsection{Where do binaries of different separations come from?}

In the following discussion we will examine the origin of the field
binary population from clustered and isolated star formation.  For
simplicity we will assume that 20 per cent of star formation occurs isolated (ISF)
\citep[actually the figure is probably somewhat less than this and could be
as low as 10 per cent;][]{Lada03}, and 40 per cent each in
low-density clusters (LDC) and high-density clusters (HDC) (i.e. 
low- and high-mass clusters with a $\beta = -2$ power-law cluster mass function).

We will also assume that the G-dwarf field separation distribution
(\citealp[i.e. the][]{Duquennoy91} Gaussian) also holds for other
stellar masses.  In particular, we assume it holds for M-dwarfs which
might not be a particularly good assumption as (a) the \citet{Fischer92} 
data is only in rough agreement with the G-dwarf
distribution, and (b) the \citeauthor{Fischer92} data probably only holds
for M-dwarfs with masses $> 0.3$\,M$_\odot$ \citep[e.g.][]{Lada06}.  However,
these assumptions at least allow us to discuss the pertinent points.

We can divide binaries into four broad categories based on their
separations.  Binaries with separations $< 50$~AU are `always hard' --
no density of cluster significantly changes the separation
distribution in this range.  Binaries with separations of $50$ --
$1000$~AU are `sometimes hard' -- high-density clusters can destroy
some of this
population, but low density clusters and isolated regions cannot.
Binaries with separations in the range $10^3$ -- $10^4$~AU are `soft-intermediates' --
high density clusters destroy such binaries, and low-density clusters
destroy some.  Binaries with separations $>10^4$~AU are `always soft'
-- any cluster will destroy such binaries (if they could even form in
the first place).

$\bullet$ Roughly 50 per cent of binaries
are `always hard'.  Such systems cannot be destroyed by all but the
most extreme cluster densities and so the field population must
represent the sum of binaries formed in HDC, LDC and ISF star
formation.

Thus, the binary fraction and separation distribution below $50$~AU
must be a fundamental outcome of the star formation process.  That is,
a combination of HDC, LDC and ISF must produce around 30 per cent of
G-dwarfs\footnote{i.e. 50 per cent of the 60 per cent of G-dwarfs in binaries.} 
and 15 -- 20 per cent of M-dwarfs with a companion $< 50$~AU
and the combined separation distribution in this range must match 
the field.

$\bullet$ Around 10 -- 15 per cent of binaries in the field are `sometimes
hard'.  Therefore star formation would be expected to slightly
over-produce such systems as many in HDCs will be destroyed, but those
in LDCs and ISF would be unaffected.

The over-production in this range need not be extreme as they survive in
around 60 per cent of star forming regions (LDCs and ISF), but up to 50
per cent of those that form may be destroyed in HDCs.  

Thus, all modes of star formation combined must produce about 10 per
cent of G-dwarfs and 5 per cent of M-dwarfs with a companion 
between $50$ and $1000$~AU.

$\bullet$ A similar fraction of 10 -- 15 per cent of binaries are
`soft-intermediate'.  Those produced in HDCs will be almost all
destroyed, many will in LDCs, but those in ISF will remain.

Following the above arguments, 40 per cent of star formation (HDCs) cannot
produce soft-intermediates, in 40 per cent of star formation half of those 
that form may be destroyed (in LDCs), and in 20 per cent (ISF) all survive.  

Thus, all modes of star formation combined must produce {\em at least}
10 per cent of G-dwarfs and 5 per cent of M-dwarfs with a companion 
between $10^3$ and $10^4$~AU.  Although the fraction formed could be
significantly larger.

$\bullet$ That leaves around 20 per cent of binaries that are `always soft' and
cannot survive (or even be formed) in any cluster.  At first
inspection, it appears that such binaries must all form in
ISF\footnote{We are investigating the possibility that they may be 
formed during the violent dissolution of clusters after gas 
expulsion \citep[][in prep]{Kouwenhoven09b}.  However, it is very unclear at the
moment if this is possible.}.

If these binaries are produced by the ISF mode, 20 per
cent of star formation must produce 20 per
cent of the total number of binaries.  This implies that almost all
isolated star formation must produce a binary with a separation
$>10^4$~AU.

However, it is difficult to see how even isolated star formation can 
produce binaries with separations $>10^4$~AU.  Isolated star forming cores only have 
radii of $\sim 0.1$~pc out to the point at which they merge with the 
background \citep[e.g.][and references therein]{Ward-Thompson07} 
and so even a companion forming at the very limit of the core from the
primary would only have a $2 \times 10^4$~AU separation (and surely disc 
fragmentation could not work at such distances). Thus the origin of 
binaries $>$ a few $\times 10^3$~AU is a mystery.

Clearly, the initial binary separation distribution cannot be identical 
to the field.  However, the form and universality (or otherwise) of the 
initial binary separation distribution remains unclear.  Note that there 
are two possible `initial' binary separation distributions (BSDs). 
Firstly, the `primordial' BSD produced as the immediate outcome of star 
formation, i.e. the distribution that emerges from Class~0/I sources. 
Secondly, the `initial' BSD that evolves rapidly from the primordial BSD 
due to circularisation and interactions with discs.  It is the 
evolution from primordial to initial BSDs that eigenevolution 
\citep{Kroupa95b} attempts to capture.

\citet{Kroupa95a,Kroupa95b} described a potential initial BSD in which 
the closer binaries ($< 50$~AU) have a field-like distribution, and wider 
binaries over-produced by a factor of $\sim 2$ out to $10^4$~AU.  However, 
as we have seen, the formation of binaries as wide as $10^4$~AU is 
problematic at best in HDCs and all will be destroyed in HDCs.  We will return 
to the problem of the form of the initial BSD and whether it is universal 
in a future paper.

\section{Conclusions}
\label{conclude}

We use $N$--body simulations to dynamically evolve star clusters rich
in binary systems to examine the effect of dynamical interactions on the
initial binary population. Our main conclusions are:

\begin{itemize}

\item{{\em Binary processing occurs within only a few initial
    crossing times}, therefore it is the initial density of a cluster
    that is of importance in determining the binary population.
    Clusters older than a few (initial) crossing times will have
    reached a dynamical equilibrium with their binary population,
    dynamically young clusters will have not.}

\item{ {\em Binaries in clusters cannot form with the field separation
  distribution.}  Binaries with separations $> 10^4$~AU are too wide
  to form in any cluster, and binaries with separations $>10^3$~AU are
  all destroyed in probably half of clusters.}

\item{ {\em Binaries in clusters cannot form with the field binary
  fraction for systems with separations in the range $50$ to a few thousand~AU.}   Many intermediate
  binaries are destroyed, therefore they must be overproduced relative
  to the levels observed in the field. }

\item{ {\em Binaries in clusters must form with roughly the correct
    binary fraction and separation distribution for systems closer
    than $\sim 50$~AU.}  Hard binaries are relatively unaffected by
    dynamical evolution; small amounts of destruction and modification
    do occur, but not enough to seriously affect this population.}

\item{ {\em The central region of the Orion Nebula Cluster was 
    initially $\sim 6$ times smaller ($\sim 200$ times denser) than it is today.  
    The binary populations were established and well-mixed at this time when the crossing time was
    significantly shorter.}  Therefore we expect there to be only
    small differences between the binary populations in the inner and
    outer regions of Orion as is observed \citep{Kohler06}. 
    However, further observations are required to help distinguish
    between different models for the current and past state of the ONC.}

\end{itemize}

A comparison of the results of star formation to the
field distribution at separations $<50$~AU probably should be made, as this
population should not normally be affected by dynamical evolution.
However, it is not clear if low- and high-density environments and
isolated star formation should always produce the same distribution(s).

It appears that star formation should -- for all masses, including
M-dwarfs -- produce more binaries with separations in the range $10^2$ -- $10^4$~AU than are
observed in the field as such binaries are highly susceptible to
destruction in clusters (depending on density).  But again, it is not 
clear if low- and high-density environments and
isolated star formation should always produce the same distribution(s).

However, star formation in clusters need not produce any binaries with separations
$> 10^4$~AU, as such binaries -- although observed in the field --
just cannot survive in clusters, even if they can form.  Their origin
is a major problem in star formation and star clusters as it cannot be
explained if most stars form in clusters.

%%%%%%%%%%%%%%%%%%%%%%%%%%%%%%%%%%%%%%%%%%%%%%%%%%%%%%%%%%%%%%%%%%%%%%
\section*{Acknowledgements}

We would like to thank the anonymous referee for their comments which lead us to
extend the scope of this work and (we hope) make it far more interesting.
RJP acknowledges financial support from STFC.  MBNK was supported by
PPARC/STFC under grant number PP/D002036/1. The authors acknowledge
the Sheffield--Bonn Royal Society International Joint Project grant,
which provided financial support and the collaborative opportunities
for this work. RJP, SPG and MBNK acknowledge the support and
hospitality  of the International Space Science Institute in Bern,
Switzerland where  part of this work was done as part of an
International Team Programme.  This work has made use of the Iceberg
computing facility, part of the White  Rose Grid computing facilities
at the University of Sheffield.

\bibliographystyle{mn2e} \bibliography{M_dwarf_ref}

\begin{thebibliography}{}

\bibitem[\protect\citeauthoryear{Aarseth, H\'enon \& Wielen}{Aarseth
  et~al.}{1974}]{Aarseth74}
Aarseth S.~J.,  H\'enon M.,    Wielen R.,  1974, A\&A, 37, 183

\bibitem[\protect\citeauthoryear{Abt, Gomez \& Levy}{Abt et~al.}{1990}]{Abt90}
Abt H.~A.,  Gomez A.~E.,    Levy S.~G.,  1990, ApJS, 74, 551

\bibitem[\protect\citeauthoryear{Allison, Goodwin, Parker, de Grijs, {Portegies
  Zwart} \& Kouwenhoven}{Allison et~al.}{2009}]{Allison09}
Allison R.~J.,  Goodwin S.~P.,  Parker R.~J.,  de Grijs R.,  {Portegies Zwart}
  S.~F.,    Kouwenhoven M. B.~N.,  2009, ApJL, submitted

\bibitem[\protect\citeauthoryear{Bastian, Gieles, Goodwin, Trancho, Smith,
  Konstantopoulos \& Efremov}{Bastian et~al.}{2008}]{Bastian08}
Bastian N.,  Gieles M.,  Goodwin S.~P.,  Trancho G.,  Smith L.~J.,
  Konstantopoulos I.,    Efremov Y.,  2008, MNRAS, 389, 223

\bibitem[\protect\citeauthoryear{Bastian \& Goodwin}{Bastian \&
  Goodwin}{2006}]{Bastian06}
Bastian N.,  Goodwin S.~P.,  2006, MNRAS, 369, L9

\bibitem[\protect\citeauthoryear{Bate}{Bate}{2009}]{Bate09}
Bate M.,  2009, MNRAS, 392, 590

\bibitem[\protect\citeauthoryear{Binney \& Tremaine}{Binney \&
  Tremaine}{1987}]{Binney87}
Binney J.,  Tremaine S.,  1987, {Galactic Dynamics}.
{Princeton, NJ, Princeton University Press, 1987, 747 p.}

\bibitem[\protect\citeauthoryear{Brandner \& K{\"o}hler}{Brandner \&
  K{\"o}hler}{1998}]{Brandner98}
Brandner W.,  K{\"o}hler R.,  1998, ApJ, 499, L79

\bibitem[\protect\citeauthoryear{Duquennoy \& Mayor}{Duquennoy \&
  Mayor}{1991}]{Duquennoy91}
Duquennoy A.,  Mayor M.,  1991, A\&A, 248, 485

\bibitem[\protect\citeauthoryear{Feigelson, Getman, Townsley, Garmire,
  Preibisch, Grosso, Montmerle, Muench \& McCaughrean}{Feigelson
  et~al.}{2005}]{Feigelson05}
Feigelson E.~D.,  Getman K.,  Townsley L.,  Garmire G.,  Preibisch T.,  Grosso
  N.,  Montmerle T.,  Muench A.,    McCaughrean M.,  2005, ApJS, 160, 379

\bibitem[\protect\citeauthoryear{Fischer \& Marcy}{Fischer \&
  Marcy}{1992}]{Fischer92}
Fischer D.~A.,  Marcy G.~W.,  1992, ApJ, 396, 178

\bibitem[\protect\citeauthoryear{Goodwin \& Bastian}{Goodwin \&
  Bastian}{2006}]{Goodwin06}
Goodwin S.~P.,  Bastian N.,  2006, MNRAS, 373, 752

\bibitem[\protect\citeauthoryear{Goodwin \& Kroupa}{Goodwin \&
  Kroupa}{2005}]{Goodwin05}
Goodwin S.~P.,  Kroupa P.,  2005, A\&A, 439, 565

\bibitem[\protect\citeauthoryear{Goodwin, Kroupa, Goodman \& Burkert}{Goodwin
  et~al.}{2007}]{Goodwin07}
Goodwin S.~P.,  Kroupa P.,  Goodman A.,    Burkert A.,  2007, in Reipurth B.,
  Jewitt D.,   Keil K.,  eds, {Protostars and Planets V} The {F}ragmentation of
  {C}ores and the {I}nitial {B}inary {P}opulation.
pp 133--147

\bibitem[\protect\citeauthoryear{Goodwin, Whitworth \& {Ward-Thompson}}{Goodwin
  et~al.}{2004}]{Goodwin04}
Goodwin S.~P.,  Whitworth A.~P.,    {Ward-Thompson} D.,  2004, A\&A, 414, 633

\bibitem[\protect\citeauthoryear{Heggie}{Heggie}{1975}]{Heggie75}
Heggie D.~C.,  1975, MNRAS, 173, 729

\bibitem[\protect\citeauthoryear{Hillenbrand \& Hartmann}{Hillenbrand \&
  Hartmann}{1998}]{Hillenbrand98}
Hillenbrand L.~A.,  Hartmann L.~W.,  1998, ApJ, 492, 540

\bibitem[\protect\citeauthoryear{Hills}{Hills}{1975a}]{Hills75a}
Hills J.~G.,  1975a, AJ, 80, 809

\bibitem[\protect\citeauthoryear{Hills}{Hills}{1975b}]{Hills75b}
Hills J.~G.,  1975b, AJ, 80, 1075

\bibitem[\protect\citeauthoryear{Jeffries}{Jeffries}{2007a}]{Jeffries07a}
Jeffries R.~D.,  2007a, MNRAS, 376, 1109

\bibitem[\protect\citeauthoryear{Jeffries}{Jeffries}{2007b}]{Jeffries07b}
Jeffries R.~D.,  2007b, MNRAS, 381, 1169

\bibitem[\protect\citeauthoryear{Jones \& Walker}{Jones \&
  Walker}{1988}]{Jones88}
Jones B.~F.,  Walker M.~F.,  1988, AJ, 95, 1755

\bibitem[\protect\citeauthoryear{K{\"o}hler}{K{\"o}hler}{2004}]{Kohler04}
K{\"o}hler R.,  2004, in Allen C.,  Scarfe S.,  eds, {Revista Mexicana de
  Astronomia y Astrofisica Conference Series} Vol.~21 of {Revista Mexicana de
  Astronomia y Astrofisica Conference Series}, {What causes the Low Binary
  Frequency in the Orion Nebula Cluster?}.
pp 104--108

\bibitem[\protect\citeauthoryear{K{\"o}hler, {Petr-Gotzens}, McCaughrean,
  Bouvier, Duch{\^e}ne, Quirrenbach \& Zinnecker}{K{\"o}hler
  et~al.}{2006}]{Kohler06}
K{\"o}hler R.,  {Petr-Gotzens} M.~G.,  McCaughrean M.~J.,  Bouvier J.,
  Duch{\^e}ne G.,  Quirrenbach A.,    Zinnecker H.,  2006, A\&A, 458, 461

\bibitem[\protect\citeauthoryear{Kouwenhoven, Brown, Goodwin, {Portegies Zwart}
  \& Kaper}{Kouwenhoven et~al.}{2009a}]{Kouwenhoven09a}
Kouwenhoven M. B.~N.,  Brown A. G.~A.,  Goodwin S.~P.,  {Portegies Zwart}
  S.~F.,    Kaper L.,  2009a, A\&A, 493, 979

\bibitem[\protect\citeauthoryear{Kouwenhoven, Goodwin, Parker \&
  Kroupa}{Kouwenhoven et~al.}{2009b}]{Kouwenhoven09b}
Kouwenhoven M. B.~N.,  Goodwin S.~P.,  Parker R.~J.,    Kroupa P.,  2009b,
  MNRAS, in prep.

\bibitem[\protect\citeauthoryear{Kouwenhoven, Brown, {Portegies Zwart} \&
  Kaper}{Kouwenhoven et~al.}{2007}]{Kouwenhoven07}
Kouwenhoven M. B.~N.,  Brown A. G.~A.,  {Portegies Zwart} S.~F.,    Kaper L.,
  2007, A\&A, 474, 77

\bibitem[\protect\citeauthoryear{Kouwenhoven, Brown, Zinnecker, Kaper \&
  {Portegies Zwart}}{Kouwenhoven et~al.}{2005}]{Kouwenhoven05}
Kouwenhoven M. B.~N.,  Brown A. G.~A.,  Zinnecker H.,  Kaper L.,    {Portegies
  Zwart} S.~F.,  2005, A\&A, 430, 137

\bibitem[\protect\citeauthoryear{Kouwenhoven \& de Grijs}{Kouwenhoven \&
  de~Grijs}{2008}]{Kouwenhoven08}
Kouwenhoven M. B.~N.,  de Grijs R.,  2008, A\&A, 480, 103

\bibitem[\protect\citeauthoryear{Kroupa}{Kroupa}{1995a}]{Kroupa95a}
Kroupa P.,  1995a, MNRAS, 277, 1491

\bibitem[\protect\citeauthoryear{Kroupa}{Kroupa}{1995b}]{Kroupa95b}
Kroupa P.,  1995b, MNRAS, 277, 1507

\bibitem[\protect\citeauthoryear{Kroupa}{Kroupa}{2000}]{Kroupa00}
Kroupa P.,  2000, New Astronomy, 4, 615

\bibitem[\protect\citeauthoryear{Kroupa}{Kroupa}{2002}]{Kroupa02}
Kroupa P.,  2002, Science, 295, 82

\bibitem[\protect\citeauthoryear{Kroupa}{Kroupa}{2008}]{Kroupa08}
Kroupa P.,  2008, in {Aarseth} S.~J.,  {Tout} C.~A.,   {Mardling} R.~A.,  eds,
  Lecture Notes in Physics, Berlin Springer Verlag Vol.~760 of Lecture Notes in
  Physics, Berlin Springer Verlag, {Initial Conditions for Star Clusters}.
p.~181

\bibitem[\protect\citeauthoryear{Kroupa, Aarseth \& Hurley}{Kroupa
  et~al.}{2001}]{Kroupa01a}
Kroupa P.,  Aarseth A.,    Hurley J.,  2001, MNRAS, 321, 699

\bibitem[\protect\citeauthoryear{Kroupa, Bouvier, Duch\^ene \& Moraux}{Kroupa
  et~al.}{2003}]{Kroupa03}
Kroupa P.,  Bouvier J.,  Duch\^ene G.,    Moraux E.,  2003, MNRAS, 346, 354

\bibitem[\protect\citeauthoryear{Kroupa \& Burkert}{Kroupa \&
  Burkert}{2001}]{Kroupa01b}
Kroupa P.,  Burkert A.,  2001, ApJ, 555, 945

\bibitem[\protect\citeauthoryear{Kroupa, Petr \& McCaughrean}{Kroupa
  et~al.}{1999}]{Kroupa99}
Kroupa P.,  Petr M.~G.,    McCaughrean M.~J.,  1999, New Astronomy, 4, 495

\bibitem[\protect\citeauthoryear{Lada}{Lada}{2006}]{Lada06}
Lada C.~J.,  2006, ApJ, 640, L63

\bibitem[\protect\citeauthoryear{Lada \& Lada}{Lada \& Lada}{2003}]{Lada03}
Lada C.~J.,  Lada E.~A.,  2003, ARA\&A, 41, 57

\bibitem[\protect\citeauthoryear{McCaughrean \& Stauffer}{McCaughrean \&
  Stauffer}{1994}]{McCaughrean94}
McCaughrean M.~J.,  Stauffer J.~R.,  1994, AJ, 108, 1382

\bibitem[\protect\citeauthoryear{Mason, Gies, Hartkopf, W.~G.~Bagnuolo, {ten
  Brummelaar} \& McAlister}{Mason et~al.}{1998}]{Mason98}
Mason B.~D.,  Gies D.~R.,  Hartkopf W.~I.,  W.~G.~Bagnuolo J.,  {ten
  Brummelaar} T.,    McAlister H.~A.,  1998, AJ, 115, 821

\bibitem[\protect\citeauthoryear{Mathieu}{Mathieu}{1994}]{Mathieu94}
Mathieu R.~D.,  1994, ARA\&A, 32, 465

\bibitem[\protect\citeauthoryear{Maxted, Jeffries, Oliveira, Naylor \&
  Jackson}{Maxted et~al.}{2008}]{Maxted08}
Maxted P. F.~L.,  Jeffries R.~D.,  Oliveira J.~M.,  Naylor T.,    Jackson
  R.~J.,  2008, MNRAS, 385, 2210

\bibitem[\protect\citeauthoryear{Mayor, Duquennoy, Halbwachs \&
  Mermilliod}{Mayor et~al.}{1992}]{Mayor92}
Mayor M.,  Duquennoy A.,  Halbwachs J.-L.,    Mermilliod J.-C.,  1992, in
  McAlister H.~A.,  Hartkopf W.~I.,  eds, {IAU Colloq. 135: Complementary
  Approaches to Double and Multiple Star Research} Vol.~32 of ASP Conference
  Series, {CORAVEL Surveys to Study Binaries of Different Masses and Ages}.
IAU, pp 73--81

\bibitem[\protect\citeauthoryear{Olczak, Pfalzner \& Eckart}{Olczak
  et~al.}{2008}]{Olczak08}
Olczak C.,  Pfalzner S.,    Eckart A.,  2008, A\&A, 488, 191

\bibitem[\protect\citeauthoryear{Patience, Ghez, Reid \& Matthews}{Patience
  et~al.}{2002}]{Patience02}
Patience J.,  Ghez A.~M.,  Reid I.~N.,    Matthews K.,  2002, AJ, 123, 1570

\bibitem[\protect\citeauthoryear{Pfalzner \& Olczak}{Pfalzner \&
  Olczak}{2007}]{Pfalzner07}
Pfalzner S.,  Olczak C.,  2007, A\&A, 475, 875

\bibitem[\protect\citeauthoryear{Plummer}{Plummer}{1911}]{Plummer11}
Plummer H.~C.,  1911, MNRAS, 71, 460

\bibitem[\protect\citeauthoryear{{Portegies Zwart}, McMillan, Hut \&
  Makino}{{Portegies Zwart} et~al.}{2001}]{Zwart01}
{Portegies Zwart} S.~F.,  McMillan S. L.~W.,  Hut P.,    Makino J.,  2001,
  MNRAS, 321, 199

\bibitem[\protect\citeauthoryear{{Portegies Zwart}, Makino, McMillan \&
  Hut}{{Portegies Zwart} et~al.}{1999}]{Zwart99}
{Portegies Zwart} S.~F.,  Makino J.,  McMillan S. L.~W.,    Hut P.,  1999,
  A\&A, 348, 117

\bibitem[\protect\citeauthoryear{Reipurth, Guimar{\~a}es, Connelley \&
  Bally}{Reipurth et~al.}{2007}]{Reipurth07}
Reipurth B.,  Guimar{\~a}es M.~M.,  Connelley M.~S.,    Bally J.,  2007, AJ,
  134, 2272

\bibitem[\protect\citeauthoryear{Scally, Clarke \& McCaughrean}{Scally
  et~al.}{1999}]{Scally99}
Scally A.,  Clarke C.,    McCaughrean M.~J.,  1999, MNRAS, 306, 253

\bibitem[\protect\citeauthoryear{Scally, Clarke \& McCaughrean}{Scally
  et~al.}{2005}]{Scally05}
Scally A.,  Clarke C.,    McCaughrean M.~J.,  2005, MNRAS, 358, 742

\bibitem[\protect\citeauthoryear{Thies \& Kroupa}{Thies \&
  Kroupa}{2008}]{Thies08}
Thies I.,  Kroupa P.,  2008, MNRAS, 390, 1200

\bibitem[\protect\citeauthoryear{Tobin, Hartmann, Furesz, Mateo \&
  Megeath}{Tobin et~al.}{2009}]{Tobin09}
Tobin J.~J.,  Hartmann L.,  Furesz G.,  Mateo M.,    Megeath S.~T.,  2009,
  {ArXiv e-prints: 0903.2775}

\bibitem[\protect\citeauthoryear{Ward-Thompson, Andr{\'e}, Crutcher, Johnstone,
  Onishi \& Wilson}{Ward-Thompson et~al.}{2007}]{Ward-Thompson07}
Ward-Thompson D.,  Andr{\'e} P.,  Crutcher R.,  Johnstone D.,  Onishi T.,
  Wilson C.,  2007, in Reipurth B.,  Jewitt D.,   Keil K.,  eds, {Protostars
  and Planets V} {An Observational Perspective of Low-Mass Dense Cores II:
  Evolution Toward the Initial Mass Function}.
pp 33--46

\end{thebibliography}

\label{lastpage}

\end{document}